\documentclass[aps,prb,amsmath,amssymb,reprint]{revtex4-2}
\usepackage{graphicx}
\usepackage{epstopdf}
\usepackage{amsmath}
\usepackage{indentfirst}
\usepackage{float}
\usepackage{subfigure}
\usepackage{bm}

\usepackage[utf8]{inputenc}
\usepackage[T1]{fontenc}

\begin{document}

\title{Quantum phase transition in a double quantum dot Josephson junction driven by electron-electron interactions}

\author{Cong Li}
\affiliation{Key Laboratory of Artificial Structures and Quantum Control (Ministry of Education), Department of Physics and 
Astronomy, Shanghai Jiaotong University, 800 Dongchuan Road, Shanghai 200240, China}

\author{Yiyan Wang}
\affiliation{Key Laboratory of Artificial Structures and Quantum Control (Ministry of Education), Department of Physics and 
Astronomy, Shanghai Jiaotong University, 800 Dongchuan Road, Shanghai 200240, China}

\author{Bing Dong}
\thanks{Author to whom correspondence should be addressed. Email:bdong@sjtu.edu.cn.}
\affiliation{Key Laboratory of Artificial Structures and Quantum Control (Ministry of Education), Department of Physics and 
Astronomy, Shanghai Jiaotong University, 800 Dongchuan Road, Shanghai 200240, China}

\begin{abstract}
In this work, we employ a surrogate BCS model with discrete energy levels to investigate a hybrid system comprising two quantum dots (QD1 and QD2), where QD1 is tunnel-coupled to two superconducting leads, and QD2 is only tunnel-coupled to QD1. Through exact diagonalization of this system, we obtain numerically exact solutions that enable rigorous computation of key physical quantities. Our analysis reveals a rich phase diagram featuring multiple controllable phase transitions mediated by quantum dot electron-electron interaction strength. Specifically, the system first undergoes an initial phase transition when tuning QD2's electron-electron interaction strength while maintaining QD1 in the non-interacting regime. Subsequent adjustment of QD1's electron-electron interaction strength induces a secondary phase transition, followed by a third transition arising from inter-dot electron-electron interaction modulation. Furthermore, we demonstrate that parallel magnetic field application can drive reversible ferromagnetic-antiferromagnetic phase transitions under specific parameter conditions. Finally, we report the emergence of non-local magnetization phenomena when subjecting QD1 to weak magnetic fields. And our results demonstrate that the orientation of nonlocal magnetization can be precisely manipulated through adjustment of the on-site interaction strength $U_2$ in QD2.
\end{abstract}
\maketitle

\section{Introduction}
The Josephson junction represents a paradigmatic mesoscopic system comprising a central weak-link region sandwiched between two superconducting electrodes. This fundamental structure has served as a versatile platform for investigating coherent quantum transport phenomena, as evidenced by extensive theoretical \cite{josephson_junction001,josephson_junction002,josephson_junction003,josephson_junction004,josephson_junction005,josephson_junction006,josephson_junction007,josephson_junction008,josephson_junction009,add1} and experimental \cite{josephson_junction_experiment001,josephson_junction_experiment002,josephson_junction_experiment003,josephson_junction_experiment004,josephson_junction_experiment005,add2}studies. A hallmark characteristic of Josephson junctions is the generation of a dissipationless supercurrent—the Josephson current—that flows without any applied voltage bias when a phase difference exists between the superconducting terminals\cite{non_bias_current001,non_bias_current002}. This equilibrium current arises from phase-coherent Andreev reflection processes at the superconductor-normal interfaces. A particularly intriguing phenomenon in these systems is the $0-\pi$ quantum phase transition \cite{QPT_interaction001,QPT_interaction002,QPT_interaction003,QPT_interaction004,QPT_interaction005,QPT_interaction006,QPT_magnetic001,QPT_magnetic002,QPT_magnetic003,QPT_magnetic004}, which manifests as a sign reversal of the Josephson current and has been unambiguously observed in experiments \cite{QPT_experiment001,QPT_experiment002,QPT_experiment003,QPT_experiment004,QPT_experiment005,QPT_experiment006,QPT_experiment007}. This transition fundamentally reflects a change in the ground state parity of the system. In quantum dot(QD)-based Josephson junctions, two distinct control parameters can drive this transition: (i) tuning the on-dot Coulomb interaction strength \cite{QPT_interaction001,QPT_interaction002,QPT_interaction003,QPT_interaction004,QPT_interaction005,QPT_interaction006}, or (ii) applying and varying an external magnetic field \cite{QPT_magnetic001,QPT_magnetic002,QPT_magnetic003,QPT_magnetic004}. Remarkably, despite their different control mechanisms, both approaches share a common microscopic origin—they induce a quantum phase transition(QPT) between two distinct ground state configurations: a singlet state (formed by Cooper pairs via Andreev bound states) and a doublet state (characterized by an unpaired electron). This transition can be understood within a unified framework of competition between Kondo screening or magnetic field and superconducting correlations, where the relative strength of these effects determines the ground state parity.

While QPTs in Josephson junctions containing a single QD in their central region are now well understood, significant research attention has shifted to double-QD systems, as evidenced by growing theoretical \cite{series001,series002,series003,series004,series005,series006,series007,parallel001,parallel002,parallel003,parallel004,parallel005,parallel006,fano001,fano002,fano003,DQDothers001,DQDothers002} and experimental\cite{DQDxperiment001,DQDxperiment002,DQDxperiment003,DQDxperiment004,DQDxperiment005,DQDxperiment006} investigations. The enhanced complexity of ground state physics in these two-QD systems gives rise to QPT phenomena that extend beyond conventional $0-\pi$ transitions, manifesting in three primary configurations: (i) the series configuration \cite{series001,series002,series003,series004,series005,series006,series007}, where interconnected QDs separately link to each superconducting lead; (ii) the parallel configuration \cite{parallel001,parallel002,parallel003,parallel004,parallel005,parallel006}, featuring both QDs jointly connected to each superconducting lead; and (iii) the Fano-type configuration \cite{fano001,fano002,fano003}, where QD1 bridges the superconductors while QD2 remains indirectly coupled. Recent work on Fano-type double interacting QD josephson junction \cite{fano002} has revealed both a second-order superconducting proximity effect and QPTs induced through QD2's electron interaction strength tuning, yet crucially omitted examination of QD1's on-site interactions and the effects of inter-dot interactions between QD1 and QD2 - either of which could drive additional QPTs and represent compelling research avenues. Furthermore, given established magnetic-field-induced QPTs in single-QD junctions, the analogous response in double-QD systems remains an open question of considerable fundamental interest, presenting an important unexplored dimension in hybrid superconducting nanostructure research.

Historically, theoretical analysis of Josephson junctions incorporating interacting QDs posed significant challenges due to the inherent complexity of treating electron correlations. Until recently, the typically effective approaches were Wilson's numerical renormalization group (NRG) \cite{NRG001,NRG002,fano002,parallel002} and quantum Monte Carlo (QMC) \cite{QMC001,QMC002,QMC003} methods, both of which are computationally demanding. The emergence of the surrogate model technique \cite{discretized001,discretized002,discretized003,discretized004,discretized005,discretized006} has provided a simpler yet rigorous alternative to these conventional methods. This innovative approach transforms the continuous leads Hamiltonian into a discrete analog through systematic discretization, enabling exact diagonalization of the complete system Hamiltonian. The resulting eigenvalues and eigenstates provide direct access to all relevant physical quantities through straightforward computation. Importantly, this method has already demonstrated its effectiveness for single-QD Josephson junctions, and its fundamental premise - the faithful representation of superconducting leads through equivalent discrete Hamiltonians - strongly suggests equal validity for double-QD systems. In the present study, we implement this approach by discretizing each superconducting lead into three equivalent sites, then employ state-space expansion techniques to perform exact diagonalization of the full Hamiltonian for a Fano-coupled double-QD Josephson junction. This methodology allows comprehensive computation of system observables, which we subsequently analyze to investigate: (i) interaction-driven QPTs under various quantum dot coupling regimes, and (ii) magnetic-field-induced phase transition phenomena in this novel nanoscale system.

The remainder of this work is structured as follows. Section II presents the complete theoretical framework, beginning with the full Hamiltonian description of our Fano-coupled double QD Josephson junction system, followed by its transformation into the equivalent surrogate model Hamiltonian through systematic discretization of the superconducting leads. In Section III, we perform exact diagonalization of the discretized Hamiltonian to obtain the complete eigenspectrum, which enables computation of key physical quantities including: (i) thermodynamic entropy, (ii) ground state parity, (iii) spin-spin correlation functions, (iv) superconducting order in the quantum dots, (v) dot occupation numbers, and (vi) Josephson supercurrent. These quantities collectively provide a comprehensive picture of the system's ground state properties and allow detailed investigation of interaction-driven QPTs. Furthermore, we examine magnetic-field-induced QPT by incorporating Zeeman coupling terms into the Hamiltonian, and analyze the emergence of non-local magnetization effects in this hybrid nanostructure. Finally, Section IV summarizes our principal findings and discusses their implications for understanding correlated electron phenomena in mesoscopic superconducting systems.

\section{Model Hamiltonian and Theoretical Method}
We theoretically investigate a Josephson junction with Fano-type coupled double QDs, as schematically shown in Fig. 1. The system's Hamiltonian can be expressed as:
\begin{equation}
H=H_{QD}+H_{SC}+H_T,
\end{equation}
with
\begin{equation}
\begin{split}
H_{QD}=&\sum_{i\sigma}\varepsilon_{i\sigma}d^{\dag}_{i\sigma}d_{i\sigma}+\sum_{i}U_{i}n_{i\uparrow}n_{i\downarrow}\\
&+U_{12}n_{1}n_{2}+t\sum_{\sigma}(d^{\dag}_{1\sigma}d_{2\sigma}+h.c.),
\end{split}
\end{equation}
\begin{equation}
H_{SC}=\sum_{\eta \bm{k} \sigma}\varepsilon_{\eta \bm{k} \sigma}c_{\eta \bm{k} \sigma}^{\dag}c_{\eta \bm{k} \sigma}+\sum_{\eta \bm{k}}(\Delta e^{i\phi_{\eta}}c_{\eta \bm{k} \uparrow}^{\dag}c_{\eta -\bm{k} \downarrow}^{\dag}+h.c.),
\end{equation}
\begin{equation}
H_{T}=\sum_{\eta \bm{k} \sigma}(V c_{\eta \bm{k} \sigma}^{\dag}d_{1\sigma}+h.c.).
\end{equation}
The operator $d^{\dag}_{i\sigma}$($d_{i\sigma}$) creates (annihilates) an electron with spin $\sigma$ and energy $\varepsilon_{i \sigma}$ on the $i$-th QD, where $U_{i}$ represents the on-site Coulomb repulsion. The interdot tunneling is characterized by the hopping amplitude $t$, while $U_{12}$ denotes the interdot Coulomb interaction, which is typically weaker than the intradot interactions($U_{12} < U_{i}$). Similarly, $c_{\eta \bm{k} \sigma}^{\dag}$($c_{\eta \bm{k} \sigma}$) creates (annihilates) an electron with spin $\sigma$, momentum $\bm{k}$, and energy $\varepsilon_{\eta \bm{k} \sigma}$ in the $\eta$-th SC lead, where $\Delta e^{i\phi_{\eta}}$ represents the complex superconducting order parameter, where $\Delta$ and $\phi$ are real numbers. The SC-QD coupling is described by the tunneling Hamiltonian $H_{T}$ with momentum-independent tunneling matrix elements $V$. We approximate the momentum summation by an energy integral using a constant density of states $\rho_{F} = \frac{1}{2D}$, where $D$ is the half-bandwidth. Throughout this work, we employ natural units with $\hbar$ = $k_B$ = $e$ = 1.

\begin{figure}[H]
\centering
\includegraphics[scale=0.5]{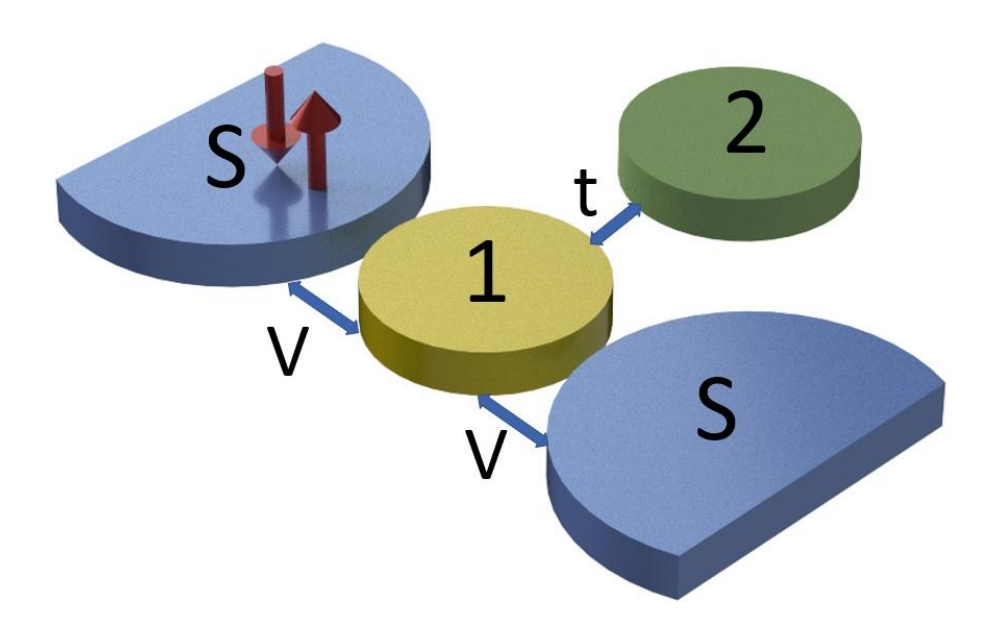}
\caption{(Color online) Schematic diagram of a QD connected to two superconductor and another QD.}
\end{figure}

Within the Nambu spinor formalism, the tunneling self-energy can be expressed as\cite{self_energy001,self_energy002,discretized006}:
\begin{equation}
	\Sigma_{T}(\omega_{n}) = -\frac{\Gamma}{4}\left[
	\begin{matrix}
		i\omega_{n} & \Delta e^{i\phi_{\eta}}\\
		\Delta e^{-i\phi_{\eta}} & i\omega_{n}
	\end{matrix}
	\right]g(\omega_{n}),
\end{equation}
with Matsubara frequencies $\omega_{n}=(2n+1)\pi T$ at temperature $T$, tunneling rate $\Gamma = 4\pi\rho_{F}V^{2}$, and the $g$ defined as
\begin{equation}
g(\omega) = \frac{2}{\pi}\frac{\arctan(\frac{D}{\sqrt{\Delta^2+\omega^2}})}{\sqrt{\Delta^2+\omega^2}}.
\end{equation}

In this paper we discretize the SC lead as three sites. The resulting $\tilde{g}$ function is given by:
\begin{equation}
\tilde{g}(\omega) = \frac{\gamma_{0}}{\xi_{0}^2+\Delta^2+\omega^2}+\frac{2\gamma_{1}}{\xi_{1}^2+\Delta^2+\omega^2},
\end{equation}
the $\tilde{g}$ function is obtained by integrating out an effective superconducting bath that has the same gap $\Delta$ as the original system. This bath consists of three discrete energy levels, located at $\xi_0$ and $\pm|\xi_1|$, which are coupled to the dot via a tunneling matrix element $\tilde{t}_l = \sqrt{\gamma_1 \Gamma}/2$. Since the number of discrete levels is odd, it follows that $\xi_0 = 0$ \cite{discretized006}. The effective bath is thus characterized by the parameters $\{{\gamma_{0}},{\gamma_{1}},\xi_{0},\xi_{1}\}$, which are determined by minimizing the cost function:
\begin{equation}
\chi^2 = \sum_{j}|g(\omega_{j})-\tilde{g}(\omega_{j})|^2,
\label{SMH1}
\end{equation}
evaluated on a nonuniform frequency grid. We employ a grid of 1000 frequency points, logarithmically spaced within the interval $\omega \in [10^{-3}\Delta, \omega_c]$, where we set $\omega_c = 10\Delta$ for all cases considered below (with $D = 1000\Delta$).

Once a satisfactory fit is obtained, the parameters $\{{\gamma_{0}},{\gamma_{1}},\xi_{0},\xi_{1}\}$ define a surrogate model Hamiltonian, representing a discretized version of the original system. This is achieved by replacing the continuous momentum variable $\bm{k}$ with a discrete index $l$, such that $\xi_{\bm{k}} \rightarrow \xi_l$ and $t \rightarrow \tilde{t}_l$. The resulting Hamiltonian takes the form:
\begin{equation}
H=H_{QD}+\tilde{H}_{SC}+\tilde{H}_T,
\label{SMH1}
\end{equation}
with
\begin{equation}
\tilde{H}_{SC}=\sum_{\eta l \sigma}\tilde{\xi}_{l}c_{\eta l \sigma}^{\dag}c_{\eta l \sigma}+\sum_{\eta l}(\Delta e^{i\phi_{\eta}}c_{\eta l \uparrow}^{\dag}c_{\eta l \downarrow}^{\dag}+h.c.),
\label{SMH2}
\end{equation}
\begin{equation}
\tilde{H}_{T}=\sum_{\eta l \sigma}(\tilde{t}_{l} c_{\eta l \sigma}^{\dag}d_{1\sigma}+h.c.),
\label{SMH3}
\end{equation}
where $c_{\eta l \sigma}^\dag$ ($c_{\eta l \sigma}$) creates (annihilates) an electron with spin $\sigma$ and energy $\xi_l$ in the $\eta$-th SC lead, characterized by the order parameter $\Delta e^{i\phi_\eta}$. The SC–QD tunneling amplitude is given by $\tilde{t}_l$. In this work, we solve the resulting low-dimensional surrogate model using exact diagonalization.
\section{Results and Discussions}
In this work, we numerically solve the low-dimensional surrogate model through exact diagonalization and subsequently calculate the system's thermodynamic properties including entropy, parity, spin correlations, QD pair term, QD occupation number, and Josephson current, with particular emphasis on the particle-hole symmetric point where $\varepsilon_{i} = -\frac{U_{i}}{2}$. Following standard conventions, we establish $\phi_{L} = \phi$ and $\phi_{R} = 0$ for the superconducting phases, with $\phi = 0$ as the default value, and maintain the system at zero temperature unless otherwise specified. Furthermore, we only focus on the strong tunneling regime ( $\Gamma$ > $\Delta$), we fix the tunneling rate at $\Gamma = 2$ and adopt the superconducting gap $\Delta = 1$ as our fundamental energy scale unless specifically stated.

\subsection{Single QD josephson junction}
We first consider the limiting case where the tunneling parameter $t$ between the two QDs is zero. In this configuration, the system reduces to a single quantum dot Josephson junction. Figure 2(a) displays the entropy as a function of the interaction strength in QD1 at various temperatures. At low temperatures, the system's entropy takes the form:
\begin{equation}
S = \ln W,
\label{entropy_state}
\end{equation}
where $S$ denotes the system entropy and $W$ represents the number of microscopic states. However, while Eq.~\eqref{entropy_state} is applicable for analyzing electronic microscopic states, the complete entropy calculation requires an alternative expression:
\begin{equation}
S = -\frac{\partial F}{\partial T},
\label{entropy_state2}
\end{equation}
where $T$ represents the temperature and $F$ denotes the system's free energy, defined as:
\begin{equation}
F = -T\ln[\sum_{n}e^{-E_{n}/T}],
\end{equation}
where $E_{n}$ represents the eigenenergy of the system, which can be obtained by solving the low-dimensional surrogate model through exact diagonalization.

\begin{figure}[h]
\centering
\includegraphics[scale=0.2]{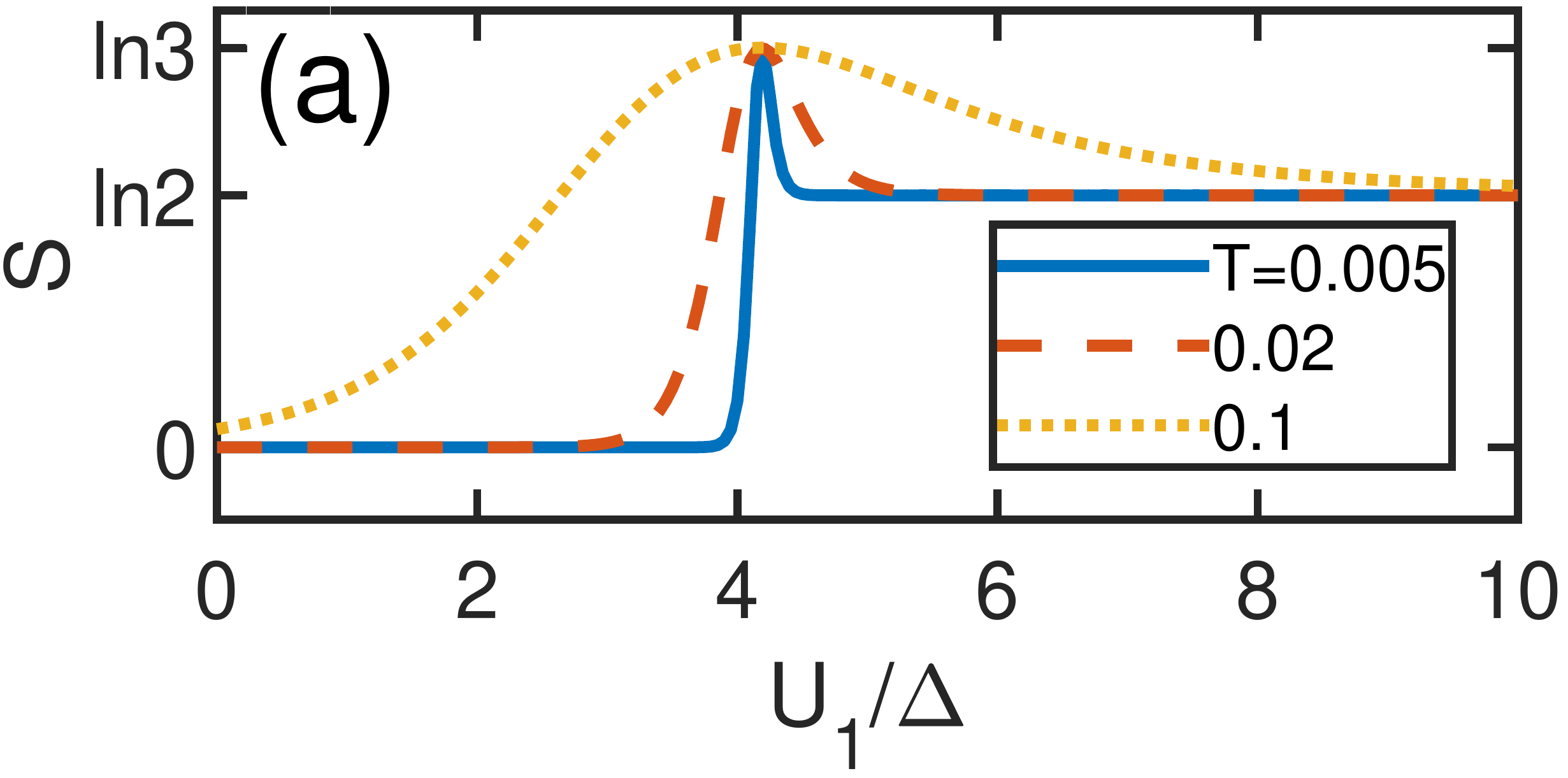}
\includegraphics[scale=0.2]{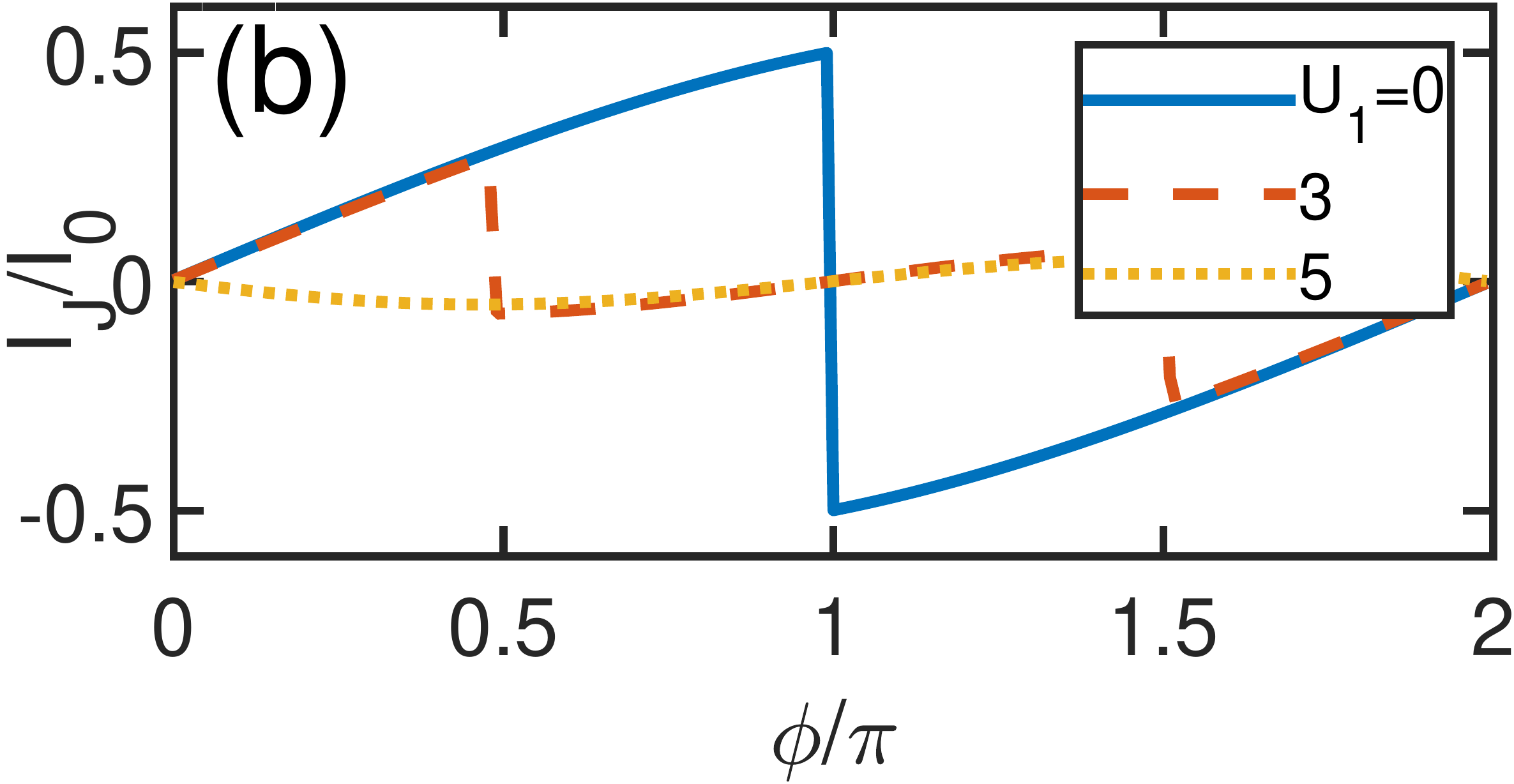}
\caption{(Colour online) (a) System entropy versus QD1 interaction strength at different temperatures $T$. (b) Josephson current as a function of the superconducting phase difference for various QD1 interaction strengths at zero temperature.}.
\end{figure}
Figure 2(a) reveals a critical point at $U_{1c} \approx 4\Delta$ for the low-temperature case ($T = 0.005$). For $U_{1} < U_{1c}$, the entropy $S=0$ as the system is in an Andreev bound states(ABS) with a singlet ground state ($W = 1$), yielding $S=\ln1=0$. When $U_{1} > U_{1c}$, the entropy becomes $S=\ln2$ due to the ground state transition to degenerate doublet states ($|\!\!\uparrow\rangle$ or $|\!\!\downarrow\rangle$, $W=2$). At $U_{1}=U_{1c}$, $S=\ln3$ occurs when the singlet and doublet states become energetically degenerate, creating three accessible ground states ($W=3$).

Figure 2(b) presents the Josephson current $I_{J}$ as a function of the superconducting phase difference for various interaction strengths in QD1 at zero temperature. The Josephson current is calculated through:
\begin{equation}
I_{J} = 2\frac{\partial E_{S}}{\partial \phi},
\label{entropy_state2}
\end{equation}
where $E_{S}$ denotes the ground state energy. The results demonstrate three distinct regimes: (i) For uncorrelated QD1, the system exhibits 0-phase behavior with positive current in the $0\to\pi$ phase difference range; (ii) When $U_{1}>U_{1c}$, the system undergoes a $0$-$\pi$ phase transition accompanied by current reversal; (iii) In the intermediate regime $0<U_{1}<U_{1c}$, the system shows $0^{'}$($\pi^{'}$)-phase characteristics. This current reversal phenomenon originates from competing contributions to the Josephson current \cite{QPT_magnetic001,QPT_magnetic002,QPT_interaction002,QPT_interaction004}: Cooper pair tunneling (positive contribution) versus single-electron tunneling (negative contribution). In the 0-phase, Cooper pair tunneling dominates, maintaining positive current for phase differences between $0$ and $\pi$. However, in the $\pi$-phase where the ground state reduces to either $|\!\!\uparrow\rangle$ or $|\!\!\downarrow\rangle$, Cooper pair tunneling is suppressed, leaving single-electron tunneling as the sole contribution and consequently reversing the current direction.

\subsection{QPT are induced by varying the electron-electron interaction parameters}
For the subsequent calculations, we fix the interdot tunneling parameter at $t = 1$ and examine the system's physical properties.
\begin{figure}[h]
\begin{minipage}{0.49\linewidth}
\centering
\includegraphics[width=1\linewidth]{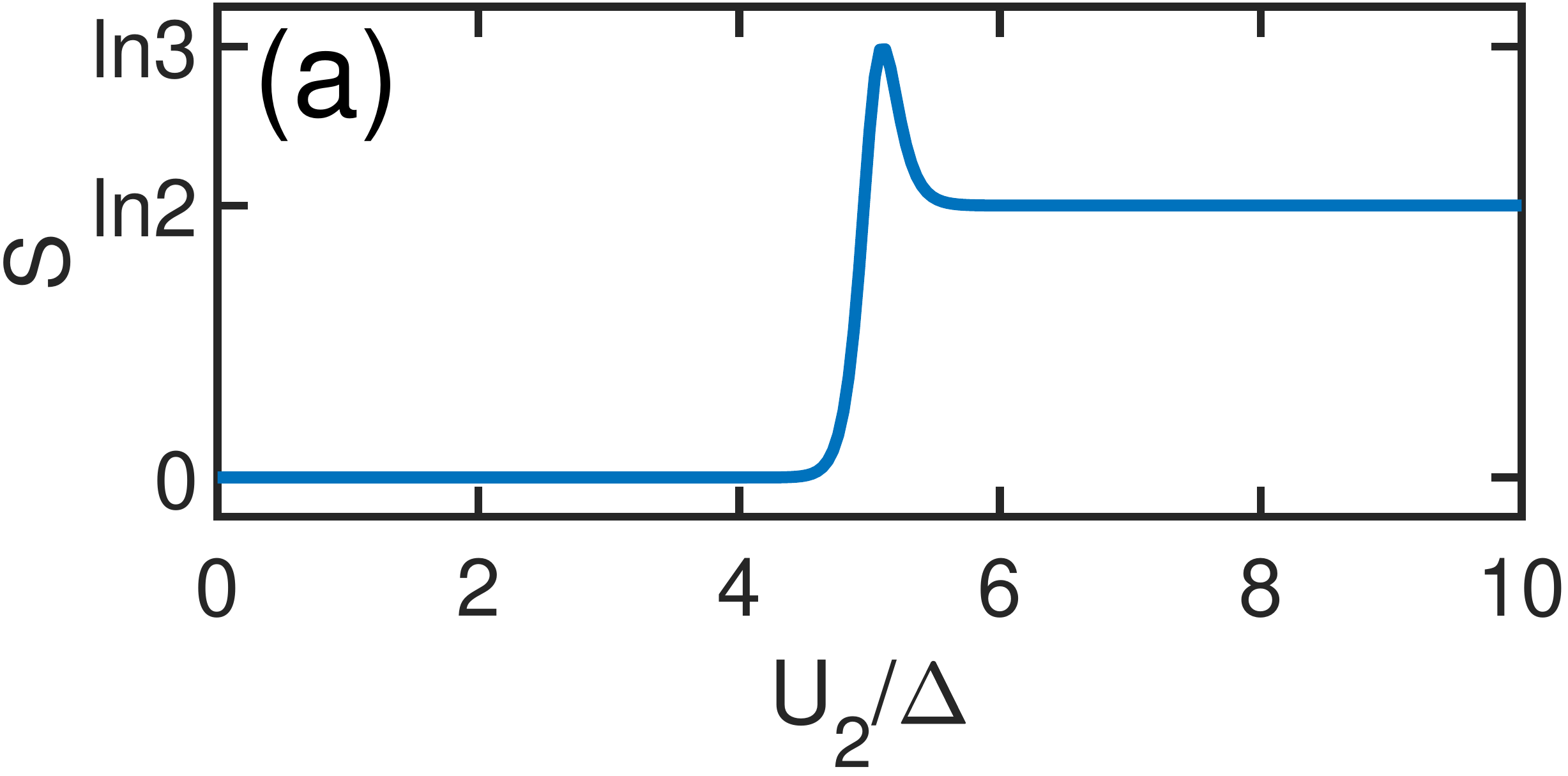}
\includegraphics[width=1\linewidth]{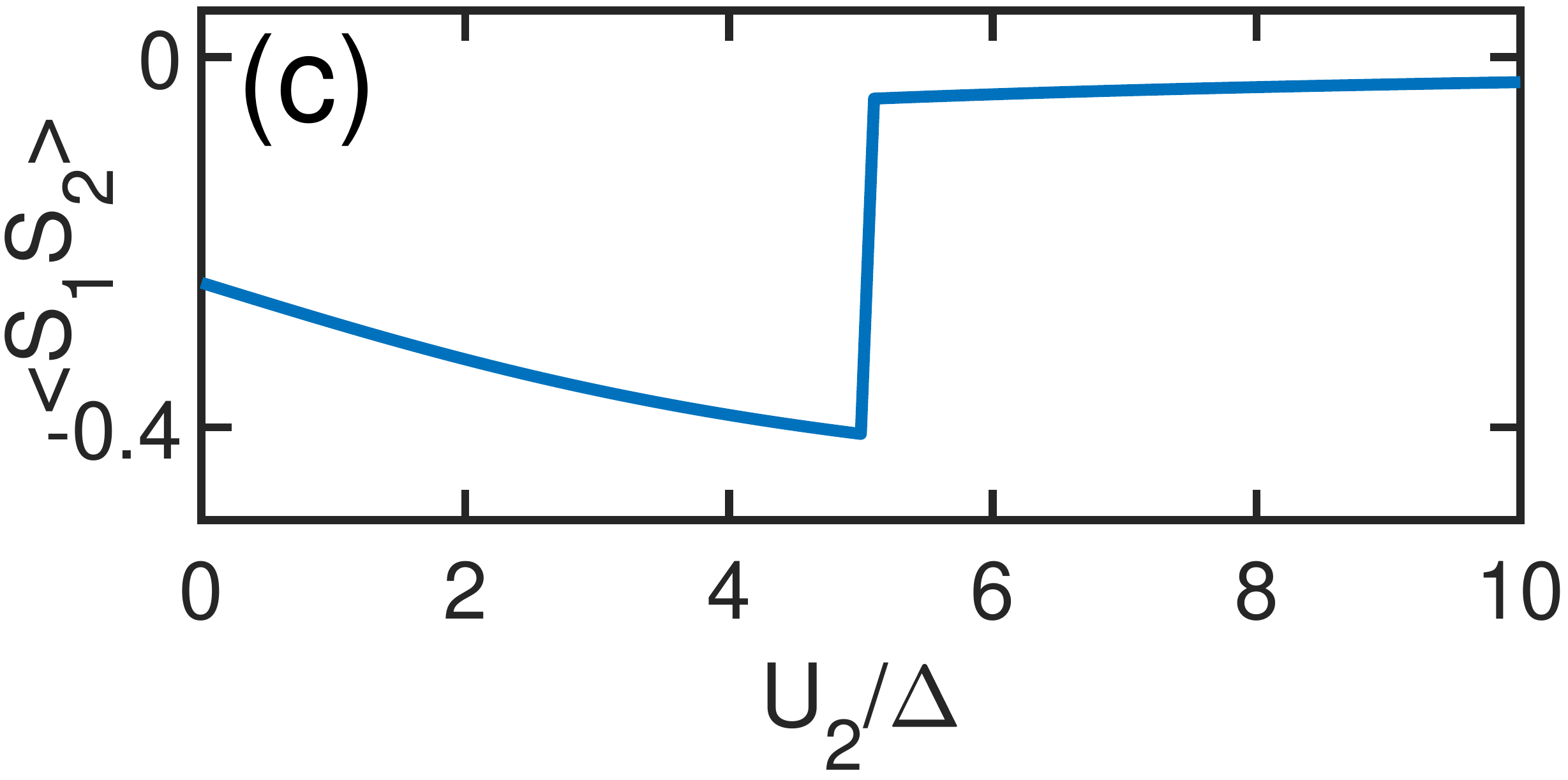}
\includegraphics[width=1\linewidth]{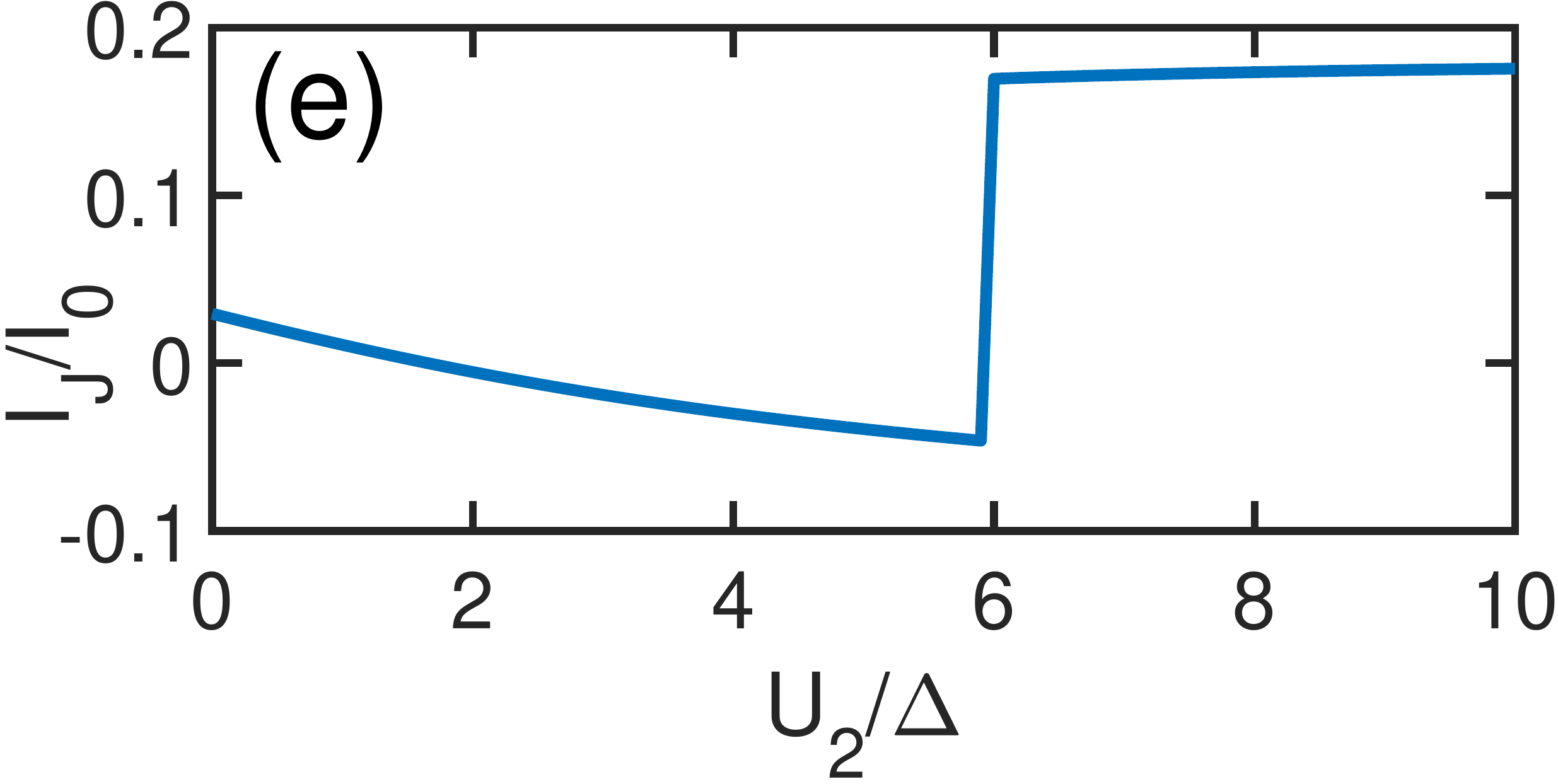}
\end{minipage}
\centering
\begin{minipage}{0.49\linewidth}
\includegraphics[width=1\linewidth]{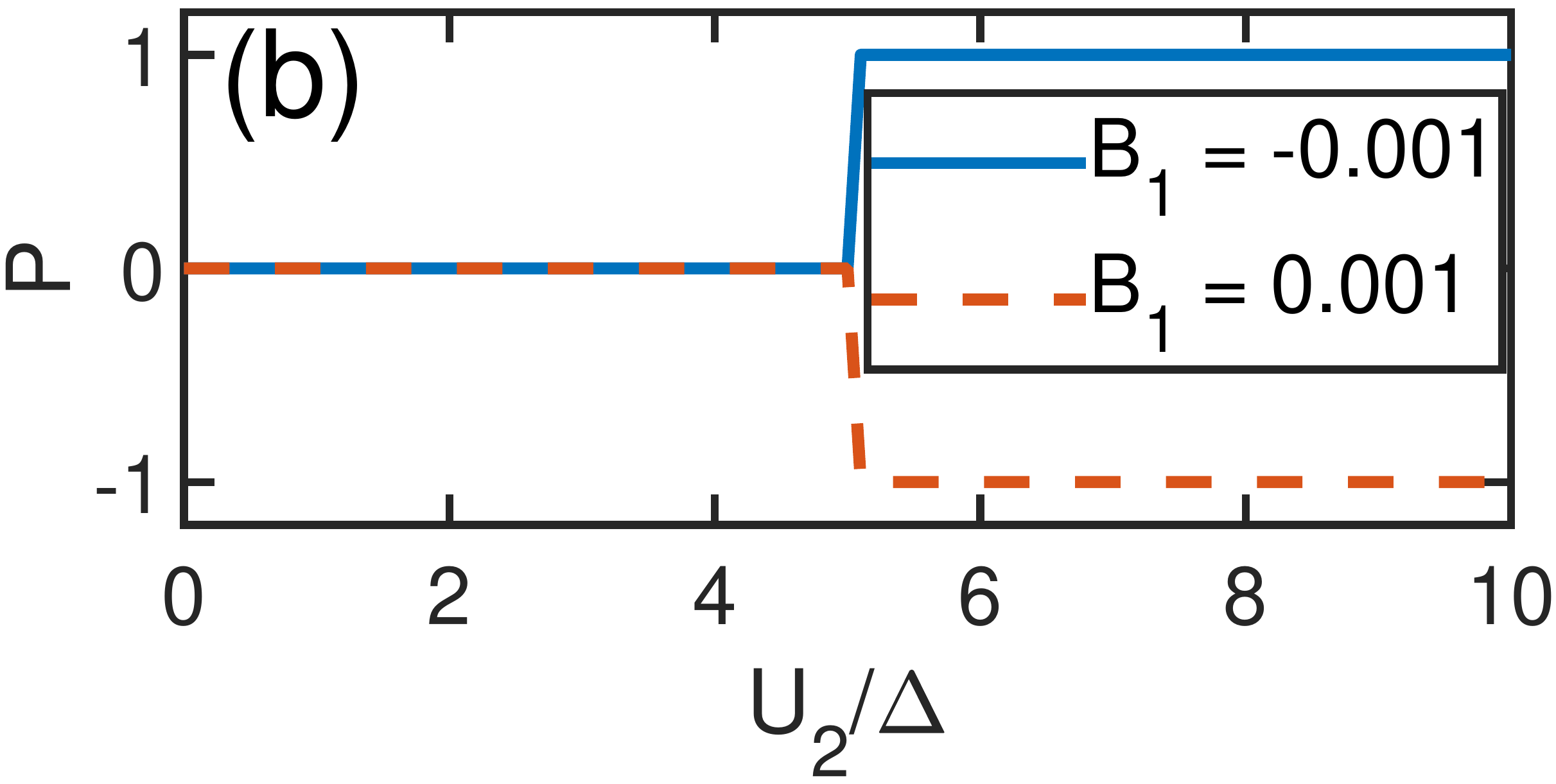}
\includegraphics[width=1\linewidth]{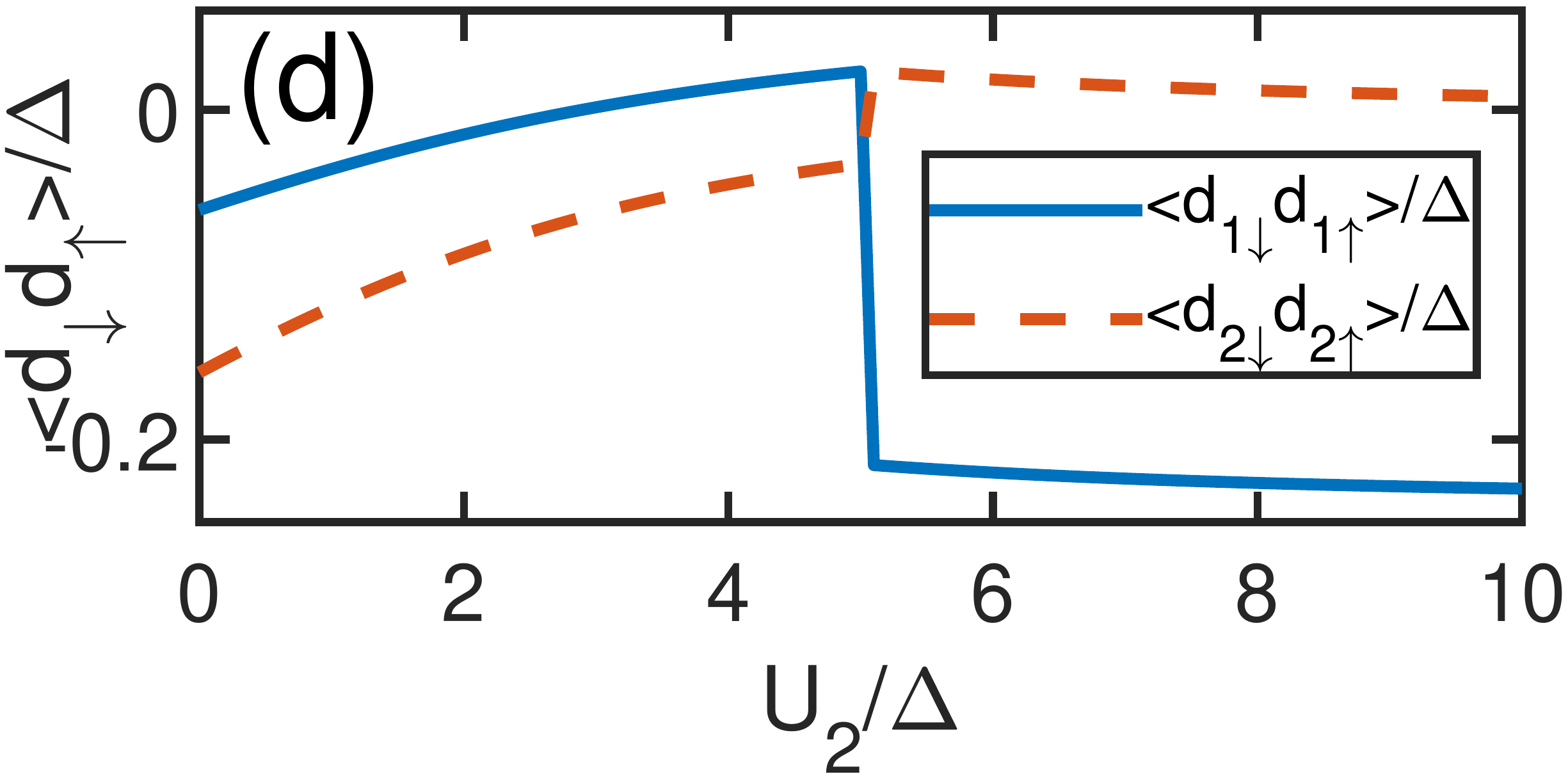}
\includegraphics[height=0.5\linewidth,width=0.97\linewidth]{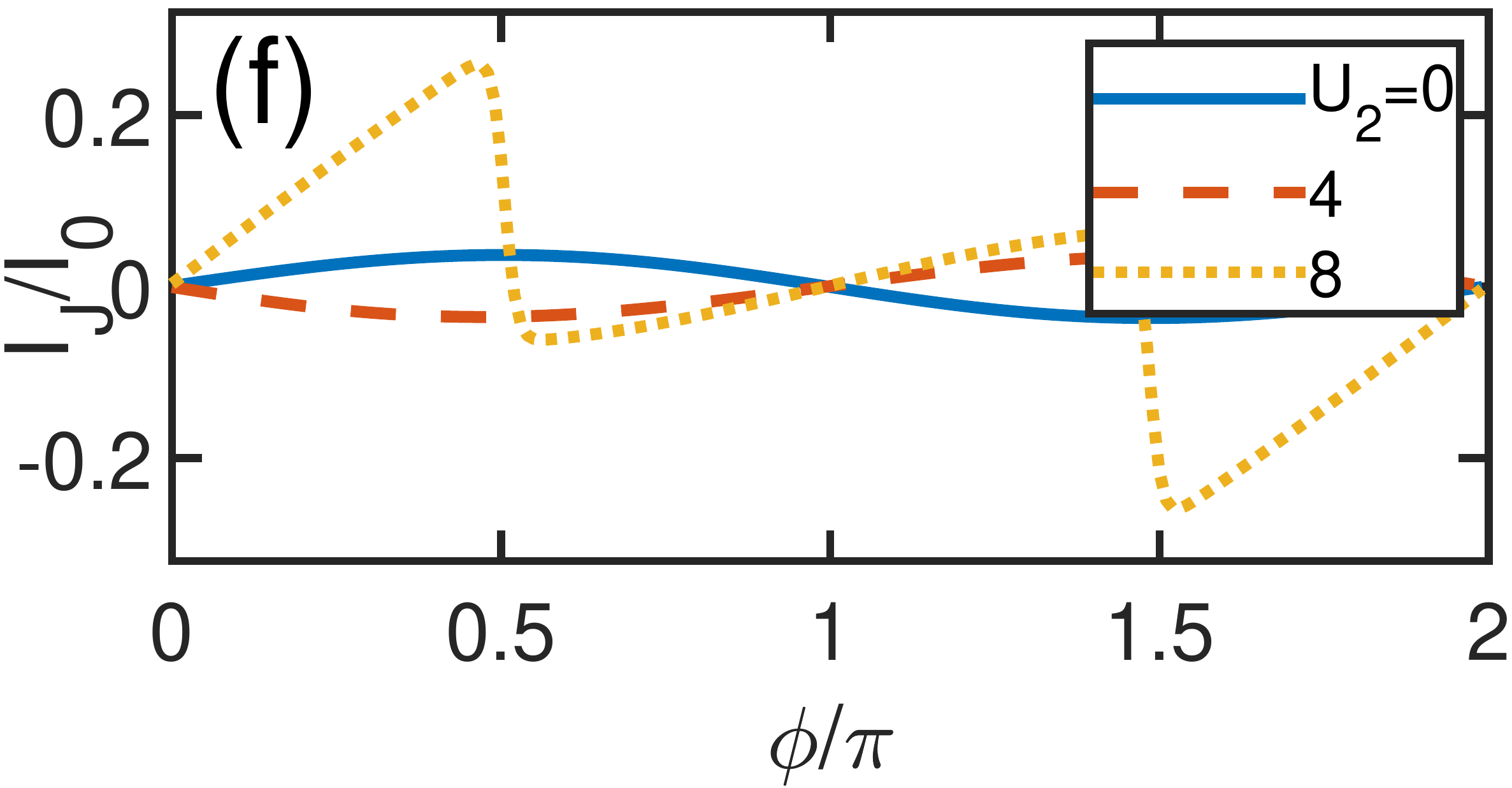}
\end{minipage}
\caption{(Colour online) System properties analysis: (a) entropy at fixed temperature $T=0.005$, (b) the expectation value of system parity operator $P$ for various magnetic field strengths $B$, (c) spin correlation function $\langle \vec{S}_{1} \!\!\cdot \!\!\vec{S}_{2} \rangle$, (d) normalized superconducting order parameters, $\langle d_{1\downarrow}d_{1\uparrow} \rangle/\Delta$ for QD1 and $\langle d_{2\downarrow}d_{2\uparrow} \rangle/\Delta$ for QD2 and (e) Josephson current $I_{J}/I_{0}$ (where $I_{0} \equiv e\Delta/\hbar$) at fixed superconducting phase difference $\phi=0.3\pi$ as a function of QD2 interaction strength $U_{2}/\Delta$. (f) $I_{J}/I_{0}$ versus phase difference $\phi$ for varying QD2 interaction strengths. The remaining system parameters are fixed at interdot tunneling $t=1$, on-site interaction $U_{1}=0$, and interdot interaction $U_{12}=0$.}.
\end{figure}

Figure 3(a) displays the entropy as a function of QD2 interaction strength $U_{2}$ at fixed temperature $T=0.005$. Throughout Fig. 3, we maintain $U_{1} = 0$ for QD1 and $U_{12} = 0$ for interdot Coulomb interaction. The entropy exhibits a QPT near $U_{2}\approx5\Delta$, with three distinct regimes: (i) $S=0$ (singlet state) for $U_{2}<5\Delta$, (ii) $S=\ln3$ at the critical point, and (iii) $S=\ln2$ (doublet state) for $U_{2}>5\Delta$. The enhanced entropy value of $\ln3$ at the quantum critical point, as previously discussed, arises from the emergent threefold degeneracy in the ground state manifold during the phase transition.
Figure 3(b) shows the parity $P$ versus $U_{2}$. Post-transition, the system enters a doublet state. To enable parity measurement, we apply a weak magnetic field to QD1, lifting the degeneracy. The parity transitions sharply from $0$ (even) to $\pm1$ (odd) at the QPT, confirming the change in ground state.
The interdot spin correlation in Fig. 3(c) plots an antiferromagnetic configuration ($\langle \vec{S}_{1} \!\cdot \!\vec{S}_{2} \rangle<0$) before the QPT that vanishes ($\langle \vec{S}_{1} \!\cdot \!\vec{S}_{2} \rangle\approx0$) in the doublet phase. The spin correlation behavior emerges from distinct ground state configurations in different phases: (i) In the singlet phase prior to the QPT, the formation of spatially localized Cooper pairs between QD1 and QD2 yields antiferromagnetic spin alignment; (ii) In the post-transition doublet phase where QD1 enters an ABS, the spin correlations become negligible.
Illustrated in Fig. 3(d) is the normalized superconducting order parameters $\langle d_{1\downarrow}d_{1\uparrow}\rangle/\Delta$ (QD1) and $\langle d_{2\downarrow}d_{2\uparrow}\rangle/\Delta$ (QD2) versus $U_{2}$. At weak interactions, both orders are significant, but they decrease with increasing $U_{2}$ due to Coulomb suppression of double occupancy and empty occupancy. After the QPT, $\langle d_{1\downarrow}d_{1\uparrow}\rangle$ surges while $\langle d_{2\downarrow}d_{2\uparrow}\rangle$ vanishes, reflecting the system's transition to a state where QD2 reaches single occupancy (particle-hole symmetry point) while QD1 forms ABS.

Through rigorous analysis of the aforementioned physical quantities, we establish that the system's ground state prior to the QPT or in the absence of electronic correlations within the QDs can be precisely described by the superposition state: $A_{1}|0_{1}0_{2}\rangle+B_{1}|0_{1}d_{2}\rangle+C_{1}|d_{1}0_{2}\rangle+D_{1}|d_{1}d_{2}\rangle+E_{1}(|\!\!\uparrow_{1}\downarrow_{2}\rangle-|\!\!\downarrow_{1}\uparrow_{2}\rangle)$, where $d_{1}$, $d_{2}$ represent the double occupation state of QD, the coefficients $A_{1}$, $B_{1}$, $C_{1}$, $D_{1}$, and $E_{1}$ quantitatively characterize the relative contributions of each entangled quantum state. Remarkably, the system undergoes a dramatic reorganization during the QPT, with the ground state abruptly collapsing into the simplified forms $|0_{1}\!\!\uparrow_{2}\rangle-|d_{1}\!\!\uparrow_{2}\rangle$ and $|0_{1}\!\!\downarrow_{2}\rangle-|d_{1}\!\!\downarrow_{2}\rangle$. This transition reveals a striking physical configuration where QD1 enters an ABS while QD2 occupies a well-defined single-particle state, yet maintains non-trivial quantum coupling with QD1. The observed discontinuities in all measured physical quantities across the phase transition boundary are fundamentally rooted in this profound reconstruction of the ground state wavefunction, demonstrating the critical relationship between quantum state reorganization and macroscopic observable phenomena.

Figure 3(e) demonstrates the Josephson current $I_J$ as a function of the Coulomb interaction strength $U_{2}$ for a superconducting phase difference $\phi = 0.3\pi$, revealing several key features of the QPT. The critical point shifts to $U_{2} \approx 6\Delta$, demonstrating the $\phi$-dependence of the transition threshold through the relation $U_{2c} = U_{2c}(\phi)$. Prior to the QPT, $I_J$ exhibits a continuous sign reversal from positive to negative with increasing $U_{2}$, this behavior emerges from the competing contributions of different ground-state components: the states $|0_{1}0_{2}\rangle$, $|0_{1}d_{2}\rangle$, $|d_{1}0_{2}\rangle$, and $|d_{1}d_{2}\rangle$ contribute to the current through two distinct mechanisms - (i) Cooper pair tunneling producing a positive current and (ii) single-electron tunneling generating a negative current, where initially the Cooper pair contribution dominates, leading to a net positive current. Since only QD1 is directly coupled to both SC leads, the antiferromagnetic state $|\!\!\uparrow_{1}\downarrow_{2}\rangle-|\!\!\downarrow_{1}\uparrow_{2}\rangle$ contributes exclusively through single-electron tunneling. This evolution is quantified by the $U_{2}$-dependent weights $A_{1}$, $B_{1}$, $C_{1}$, $D_{1} \rightarrow 0$ and $E_{1} \rightarrow 1$, causing the net current to reverse sign. Across the QPT, a second current reversal occurs from negative ($I_J < 0$ for $U_{2} \lesssim 6\Delta$) to positive values ($I_J > 0$ for $U_{2} \gtrsim 6\Delta$) as QD1 enters an ABS region , which strongly enhances Cooper pair transport in the strong correlation regime.

In Fig. 3(f), we systematically investigate the Josephson current $I_J$ as a function of the superconducting phase difference $\phi$ for various $U_{2}$. For the non-interacting case ($U_{2}=0$), the current maintains a positive sign throughout the entire phase range $0 \leq \phi \leq \pi$, exhibiting conventional Josephson behavior. At intermediate interaction strength ($U_{2} = 4\Delta$), we observe the predicted current reversal phenomenon discussed earlier. Most remarkably, in the strongly correlated regime ($U_{2} = 8\Delta$), the current-phase relation qualitatively mimics that of a single quantum dot Josephson junction in the $0^{'}(\pi^{'})$ phase, a direct consequence of the $\phi$-dependent critical interaction strength $U_{2c}(\phi)$ required to drive the quantum phase transition. This $\phi$-dependence of the transition threshold fundamentally alters the system's transport characteristics, enabling the observed rich variety of current-phase relations across different interaction regimes.

\begin{figure}[h]
\begin{minipage}{0.49\linewidth}
\centering
\includegraphics[width=1\linewidth]{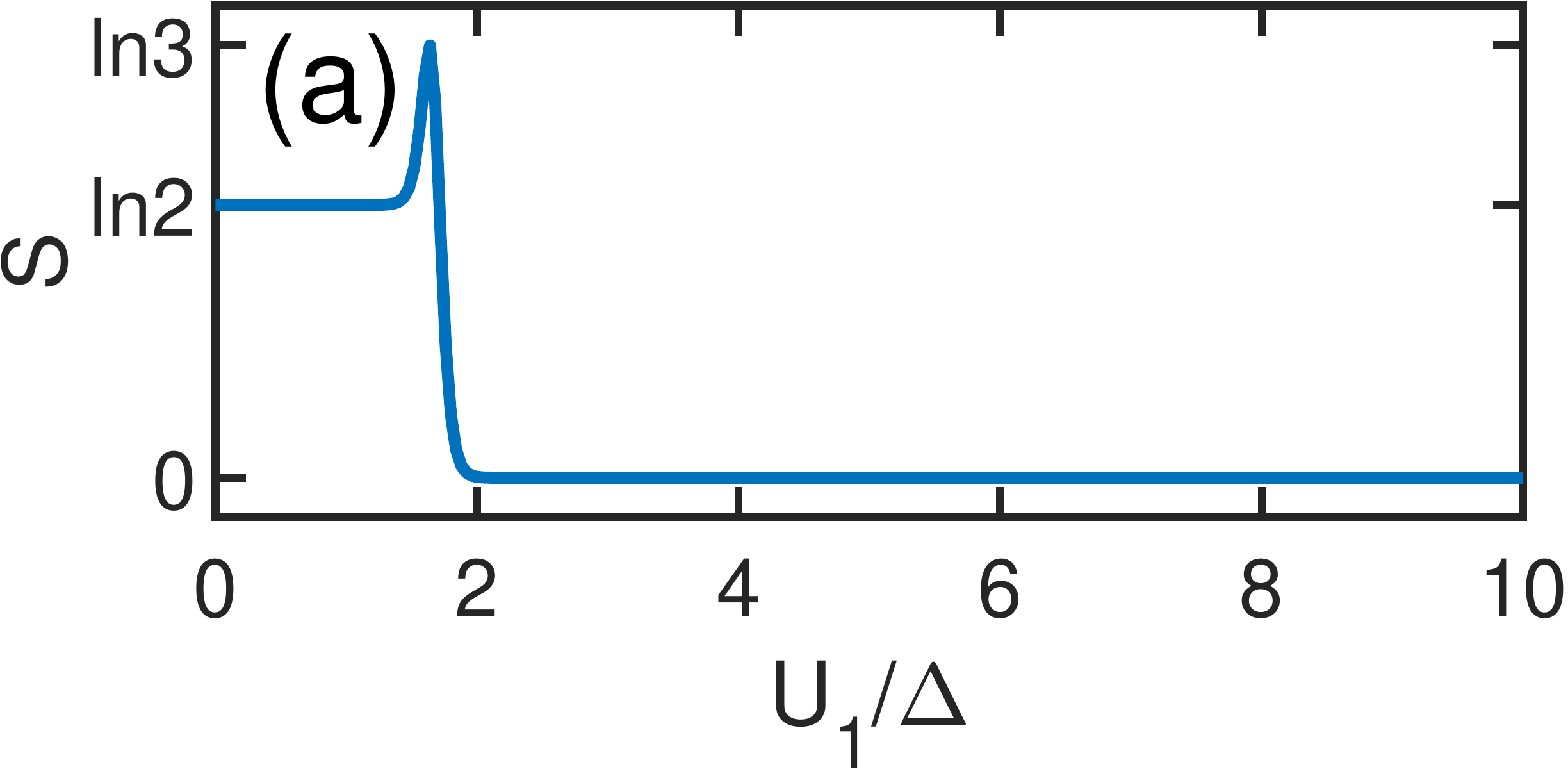}
\includegraphics[width=1\linewidth]{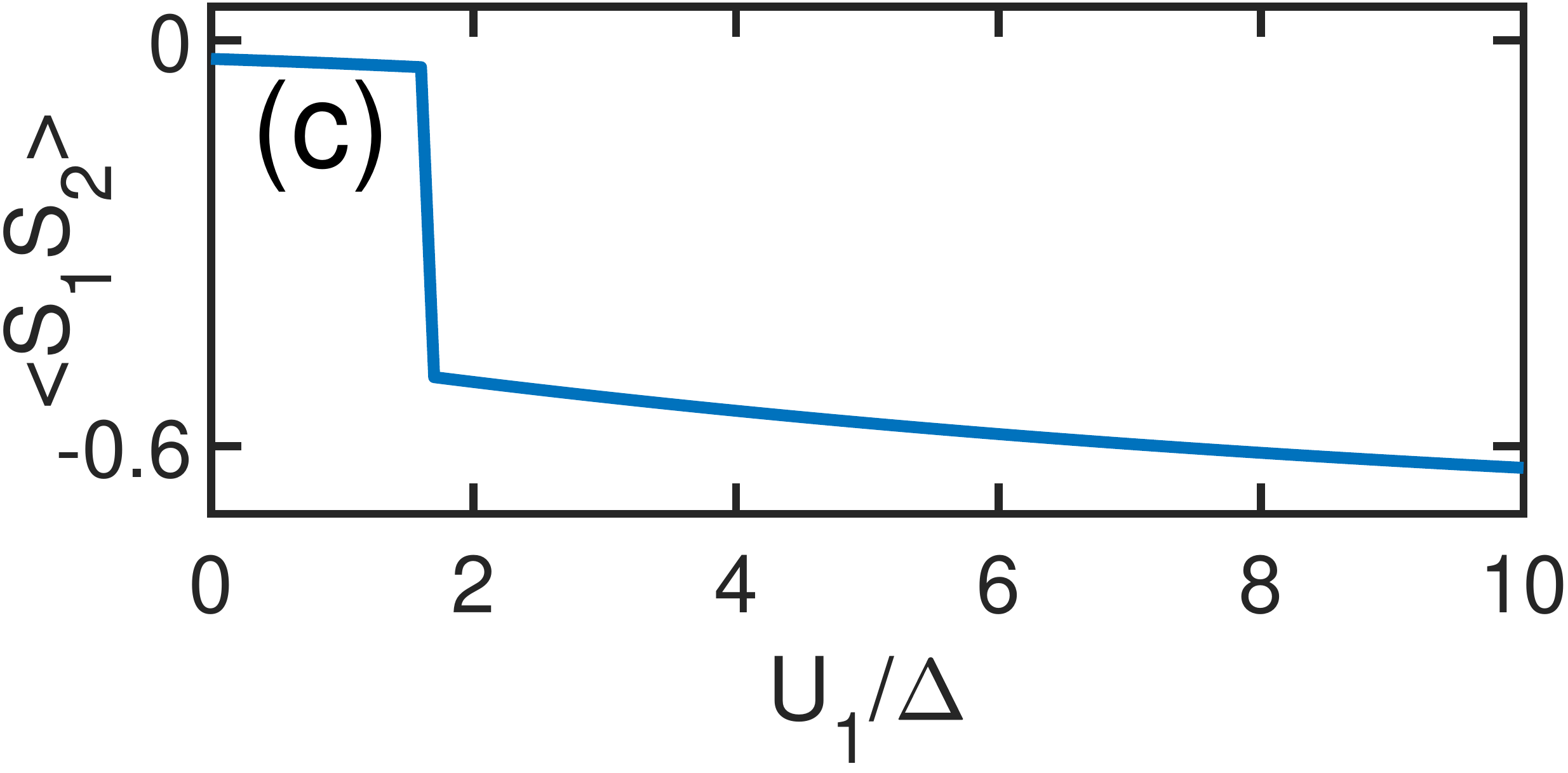}
\includegraphics[width=1\linewidth]{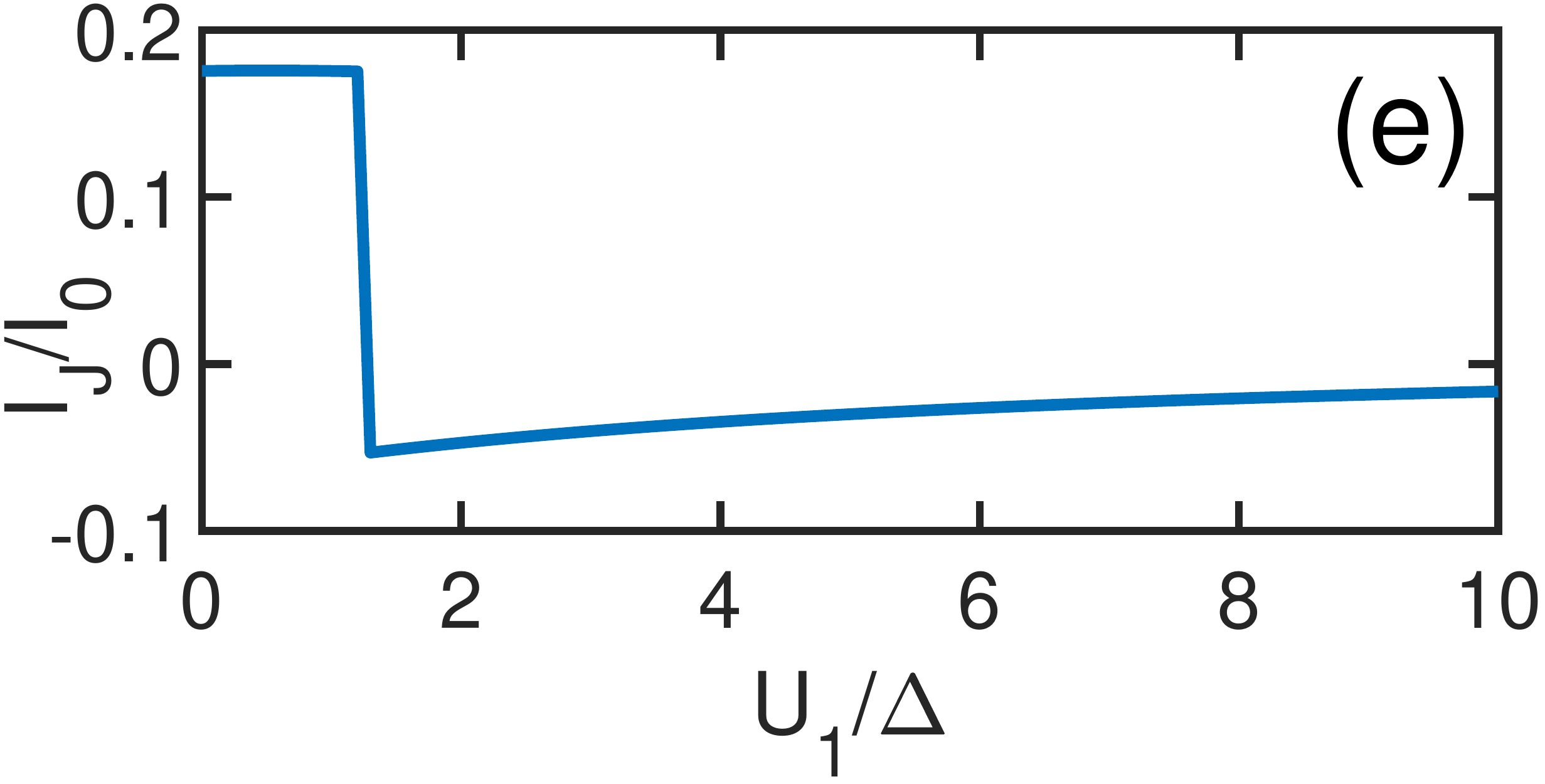}
\end{minipage}
\centering
\begin{minipage}{0.49\linewidth}
\includegraphics[width=1\linewidth]{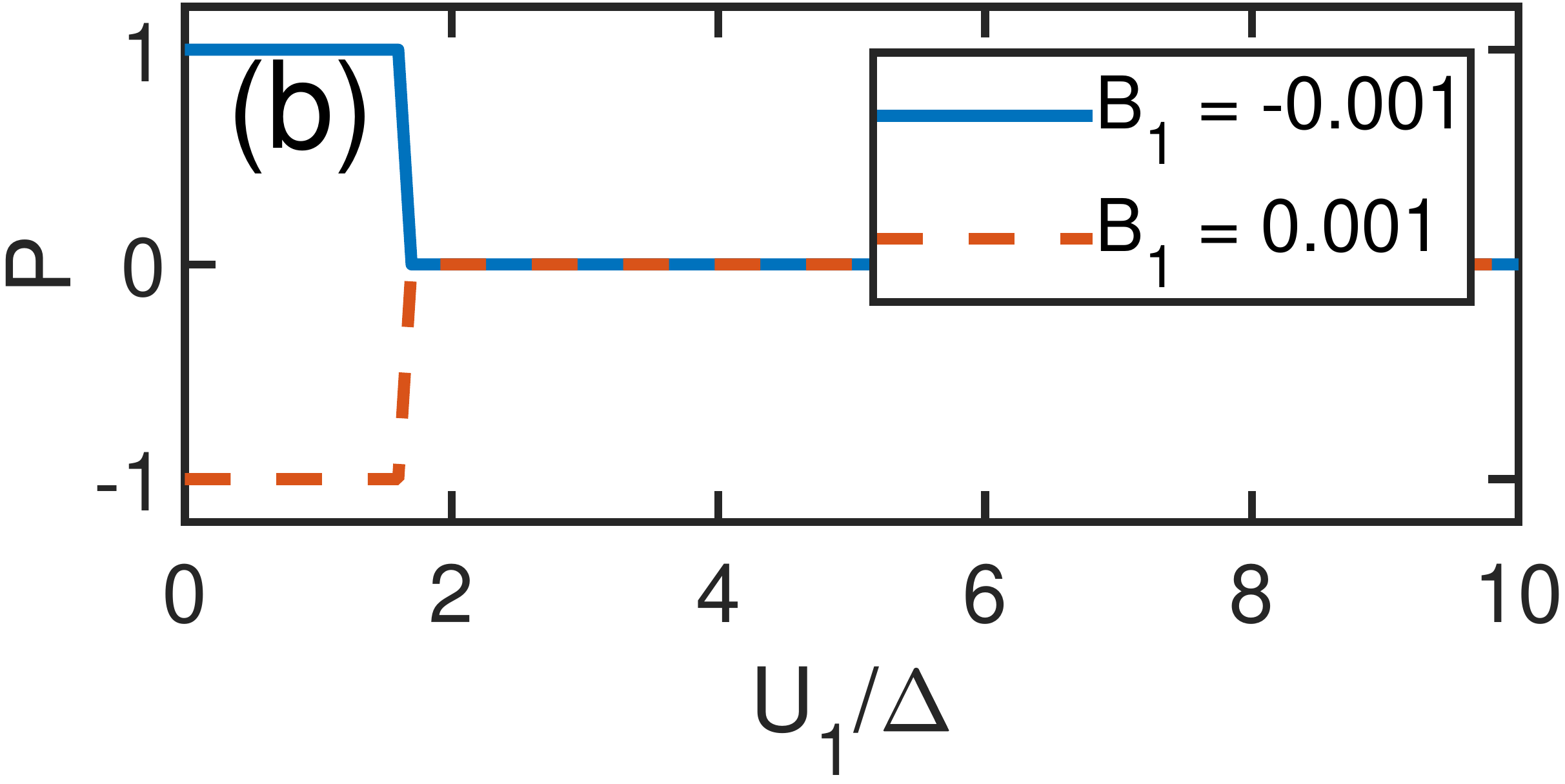}
\includegraphics[width=1\linewidth]{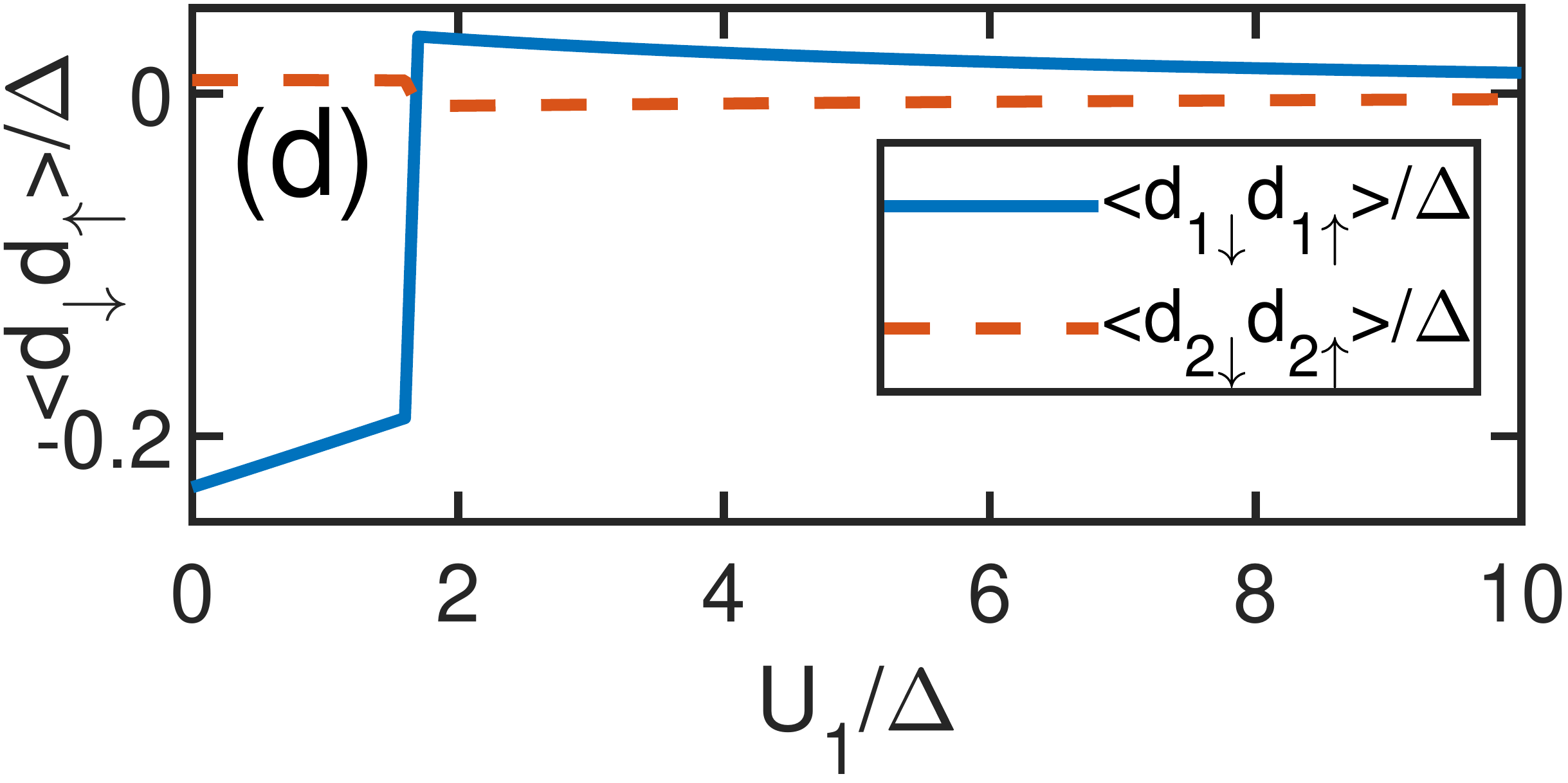}
\includegraphics[width=0.98\linewidth,height=0.5\linewidth]{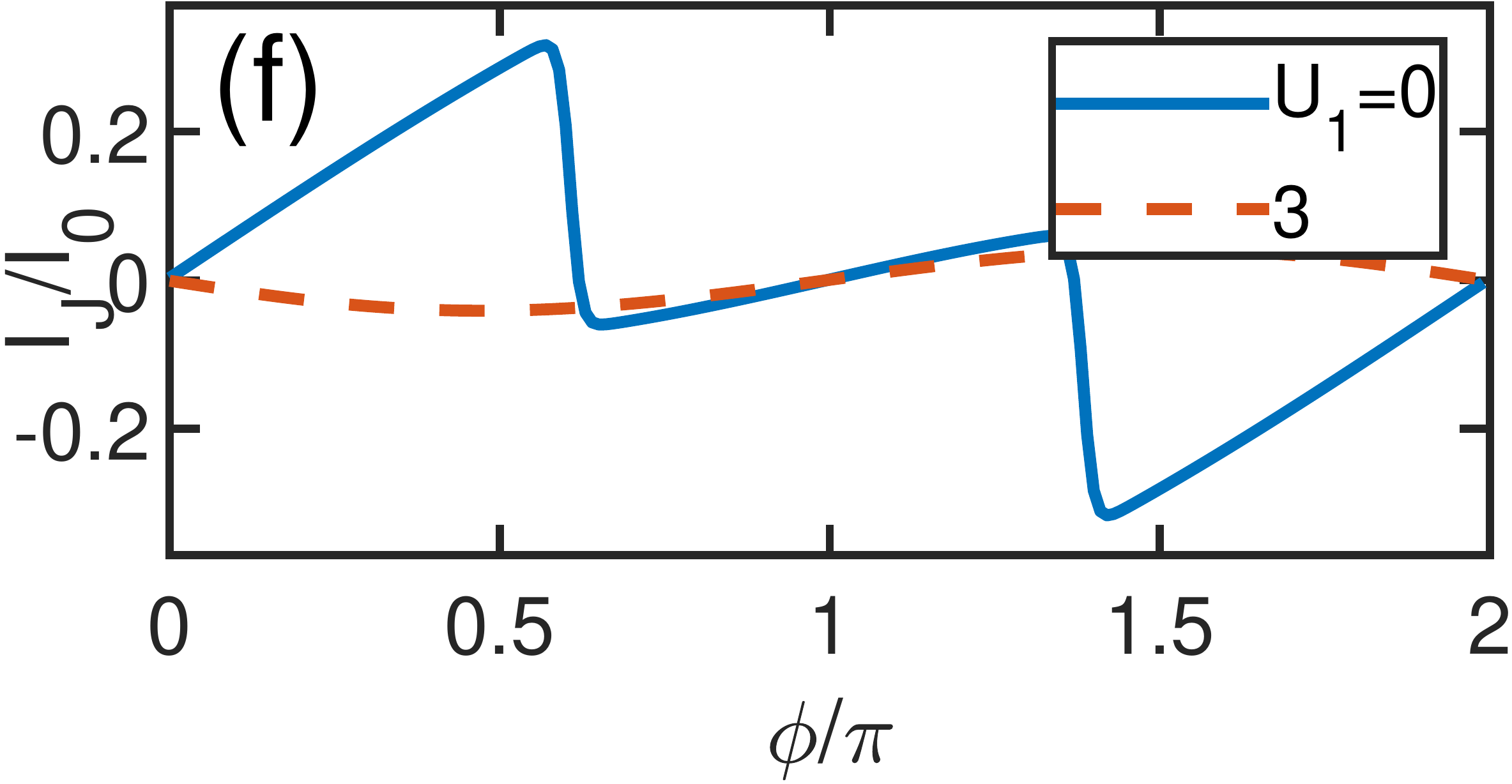}
\end{minipage}
\caption{(Colour online) System properties analysis: (a) entropy at fixed temperature $T=0.005$, (b) $P$ for various magnetic field strength $B$, (c) $\langle \vec{S}_{1} \!\!\cdot \!\!\vec{S}_{2} \rangle$, (d) $\langle d_{1\downarrow}d_{1\uparrow} \rangle/\Delta$ and $\langle d_{2\downarrow}d_{2\uparrow} \rangle/\Delta$ and (e) $I_{J}/I_{0}$ (where $I_{0} \equiv e\Delta/\hbar$) at $\phi=0.3\pi$ as a function of QD1 interaction strength $U_{1}/\Delta$. (f) $I_{J}/I_{0}$ versus phase difference $\phi$ for varying QD1 interaction strengths. While keeping the interaction strength in QD2 fixed at $U_2 = 10\Delta$ and maintaining all other parameters identical to those used in Fig. 3 unless otherwise specified.}.
\end{figure}

Figure 4(a) displays the entropy $S$ as a function of QD1's interaction strength $U_{1}$ at a low temperature $T = 0.005$, with fixed parameters $U_{2} = 10\Delta$ and $U_{12} = 0$ for QD2 and interdot interaction respectively. The entropy undergoes a sharp transition at $U_{1} \approx 1.6\Delta$, where it drops from $S = \ln2$ (indicating two-fold degenerate ground states) to $S = 0$ (signaling a non-degenerate singlet ground state), clearly marking the QPT. This transition is further characterized in Fig. 4(b) through the parity $P$, which changes from odd ($P = \pm 1$) in the pre-QPT phase to even ($P = 0$) in the post-QPT phase. The spin-spin correlation between dots, shown in Fig. 4(c), reveals the formation of an antiferromagnetic state after the QPT, while remaining nearly uncorrelated before the transition. Finally, Fig. 4(d) demonstrates the superconducting order parameters $\langle d_{1\downarrow}d_{1\uparrow}\rangle$ and $\langle d_{2\downarrow}d_{2\uparrow}\rangle$ as functions of $U_{1}$: in the pre-QPT phase ($U_{1} < 1.6\Delta$), QD1 maintains significant pairing ($|\langle d_{1\downarrow}d_{1\uparrow}\rangle| \gg 0$) characteristic of an ABS while QD2 shows negligible pairing ($\langle d_{2\downarrow}d_{2\uparrow}\rangle \approx 0$) corresponding to a single-electron state; post-QPT ($U_{1} > 1.6\Delta$), both order parameters vanish ($\langle d_{1\downarrow}d_{1\uparrow}\rangle, \langle d_{2\downarrow}d_{2\uparrow}\rangle \rightarrow 0$), indicating the complete suppression of superconductivity and transition to single-electron states in both quantum dots. Remarkably, the two quantum dots establish a non-local Cooper pairing under these specific conditions.

For Fig. 4, our analysis of the pre-transition ground state has been comprehensively presented, while the post-transition ground state can be unambiguously determined from the previously examined physical quantities to be the antiferromagnetic singlet state $|\!\!\uparrow_{1}\downarrow_{2}\rangle-|\!\!\downarrow_{1}\uparrow_{2}\rangle$, which exhibits zero parity ($P=0$) and features both QD1 and QD2 in single-particle states yet quantum-mechanically coupled through the formation of an interdot Cooper pair, as evidenced by the finite spin-spin correlation $\langle \vec{S}_{1} \!\!\cdot \vec{S}_{2} \rangle < 0$ and the complete suppression of local superconducting order parameters $\langle d_{i\downarrow}d_{i\uparrow} \rangle = 0$ ($i=1,2$) in the post-transition regime. At the particle-hole symmetric point, the Coulomb interaction effectively suppresses both doubly occupied and empty states, thereby enforcing QD1 and QD2 into single-electron configurations; however, the superconducting proximity effect mediates the formation of a nonlocal Cooper pair between the quantum dots, resulting in the post-QPT ground state being characterized by the antiferromagnetic singlet state $(|\!\!\uparrow_{1}\downarrow_{2}\rangle-|\!\!\downarrow_{1}\uparrow_{2}\rangle)/\sqrt{2}$ that combines single-electron localization with nonlocal superconducting correlations.

In Fig. 4(e), we systematically investigate the Josephson current $I_J$ as a function of the Coulomb interaction strength $U_1$ for a fixed superconducting phase difference $\phi=0.3\pi$, where we observe a distinct discontinuity in $I_J$ near the QPT point accompanied by a sign reversal - transitioning from positive current ($I_J>0$) in the pre-QPT regime to negative current ($I_J<0$) in the post-QPT regime. This behavior stems from QD1's transition from an ABS that supports positive Cooper pair current to a single-electron state that permits only negative single-particle tunneling current, despite the formation of an interdot antiferromagnetic pair between QD1 and QD2, because the superconducting leads couple exclusively to QD1, making $\phi=0.3\pi$ yield negative current in the post-QPT regime.

In Fig. 4(f), we examine the Josephson current $I_J$ as a function of the superconducting phase difference $\phi$ for varying interaction strengths $U_1$, where we observe distinct current-phase relations that correspond to different quantum states: for the non-interacting case ($U_1 = 0$), the current exhibits characteristic $0'(\pi')$-phase behavior, while at intermediate interaction strength ($U_1 = 3\Delta$), the current-phase relation undergoes a $\pi$-shift that reflects the system's transition to a $\pi$-phase state, a direct consequence of QD1 entering a single-electron configuration where the superconducting correlations vanish ($\langle d_{1\downarrow}d_{1\uparrow}\rangle = 0$) and single-particle tunneling processes dominate the transport.

\begin{figure}[h]
\centering
\includegraphics[scale=0.2]{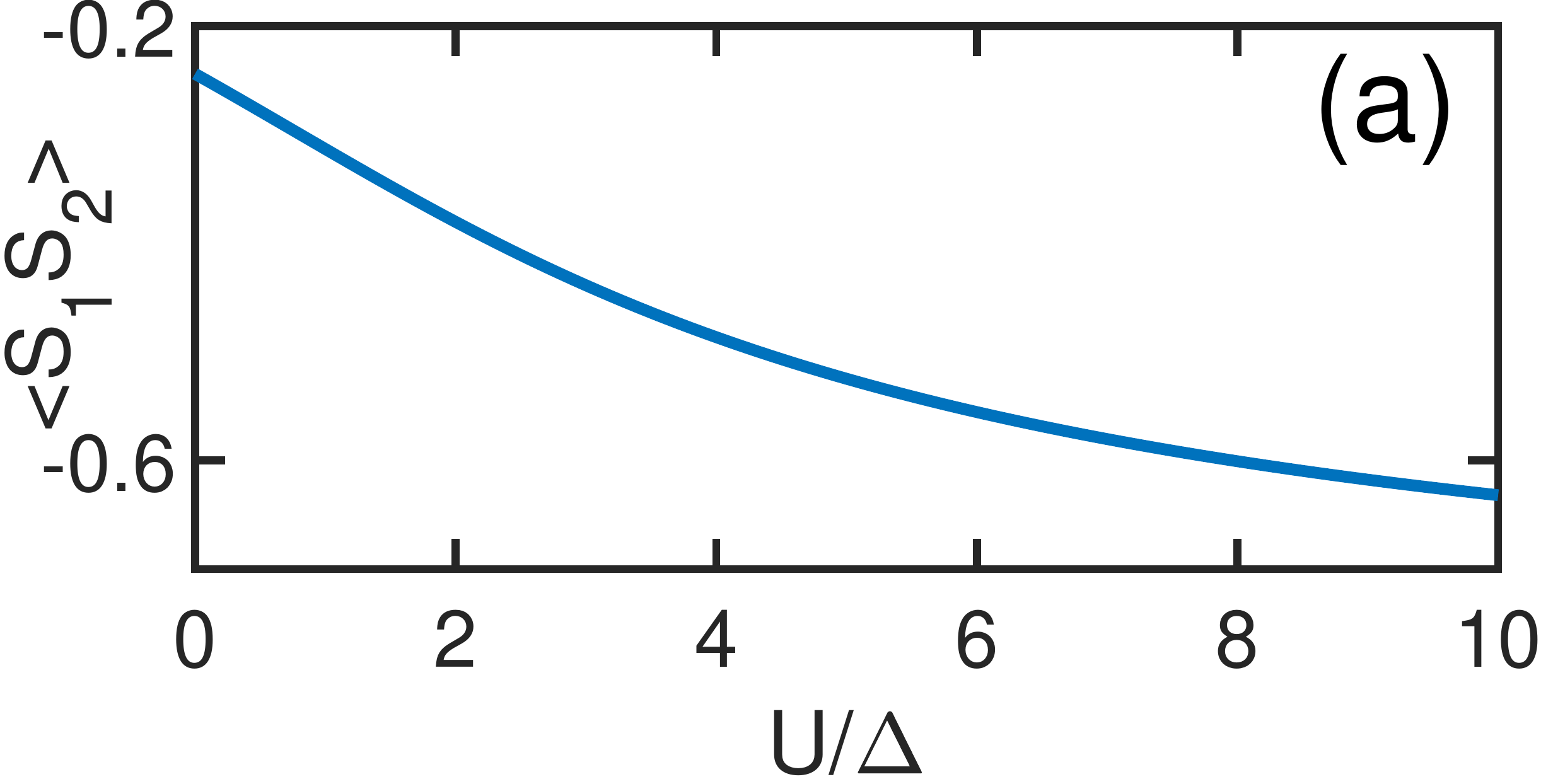}
\includegraphics[scale=0.2]{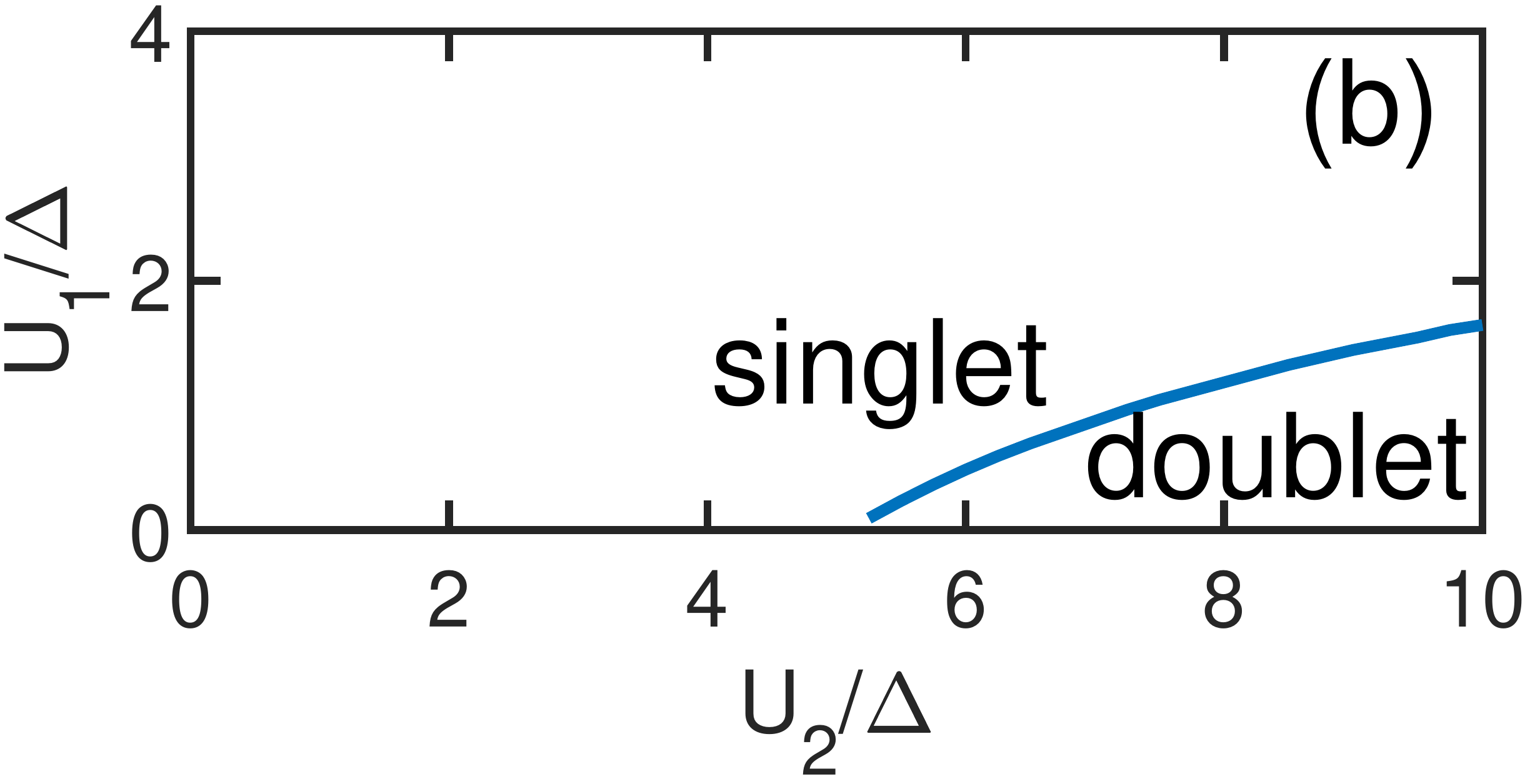}
\caption{(Colour online) (a) The interdot spin correlation $\langle \vec{S}_{1} \!\!\cdot \!\!\vec{S}_{2} \rangle$ as a function of the uniform Coulomb interaction strength $U = U_{1} = U_{2}$ in both quantum dots. (b) Phase diagram characterizing the singlet-doublet state transition as a function of the normalized interaction parameters $U_{1}/\Delta$ and $U_{2}/\Delta$, with all other system parameters maintained identical to those specified in Fig. 3.}.
\end{figure}

Our previous calculations first introduced electron-electron interaction specifically on QD2, revealing a QPT, followed by introducing interaction on QD1 while maintaining finite interaction on QD2, which induced another QPT; nevertheless, the fundamental question of whether similar phase transitions would emerge when interactions are introduced first on QD1 or simultaneously on both QDs remained unresolved, prompting the systematic investigation presented in Fig. 5(a) where we set $U_{1}=U_{2}=U$ and examined the spin correlation function $\langle \vec{S}_{1} \!\!\cdot\!\! \vec{S}_{2} \rangle$ as a continuous function of $U$, finding smooth evolution without any non-analytic behavior that would signal a QPT - in this symmetric interaction case, the ground state evolves smoothly from the non-interacting configuration $A_{1}|0_{1}0_{2}\rangle+B_{1}|0_{1}d_{2}\rangle+C_{1}|d_{1}0_{2}\rangle+D_{1}|d_{1}d_{2}\rangle+E_{1}(|\!\!\uparrow_{1}\downarrow_{2}\rangle-|\!\!\downarrow_{1}\uparrow_{2}\rangle)$ to the strongly correlated limit $(|\!\!\uparrow_{1}\downarrow_{2}\rangle-|\!\!\downarrow_{1}\uparrow_{2}\rangle)/\sqrt{2}$, as the increasing Coulomb interaction $U$ suppresses both doubly occupied and empty states at the particle-hole symmetric point, causing the coefficients $A_{1}$, $B_{1}$, $C_{1}$ and $D_{1}$ to diminish while $E_{1}$ remains dominant, with this comprehensive picture being further corroborated by the phase diagram in Fig. 5(b) that maps the singlet-doublet phase boundary across the full parameter space of $U_{1}$ and $U_{2}$ interactions.

\begin{figure}[h]
\begin{minipage}{0.49\linewidth}
\centering
\includegraphics[width=1\linewidth]{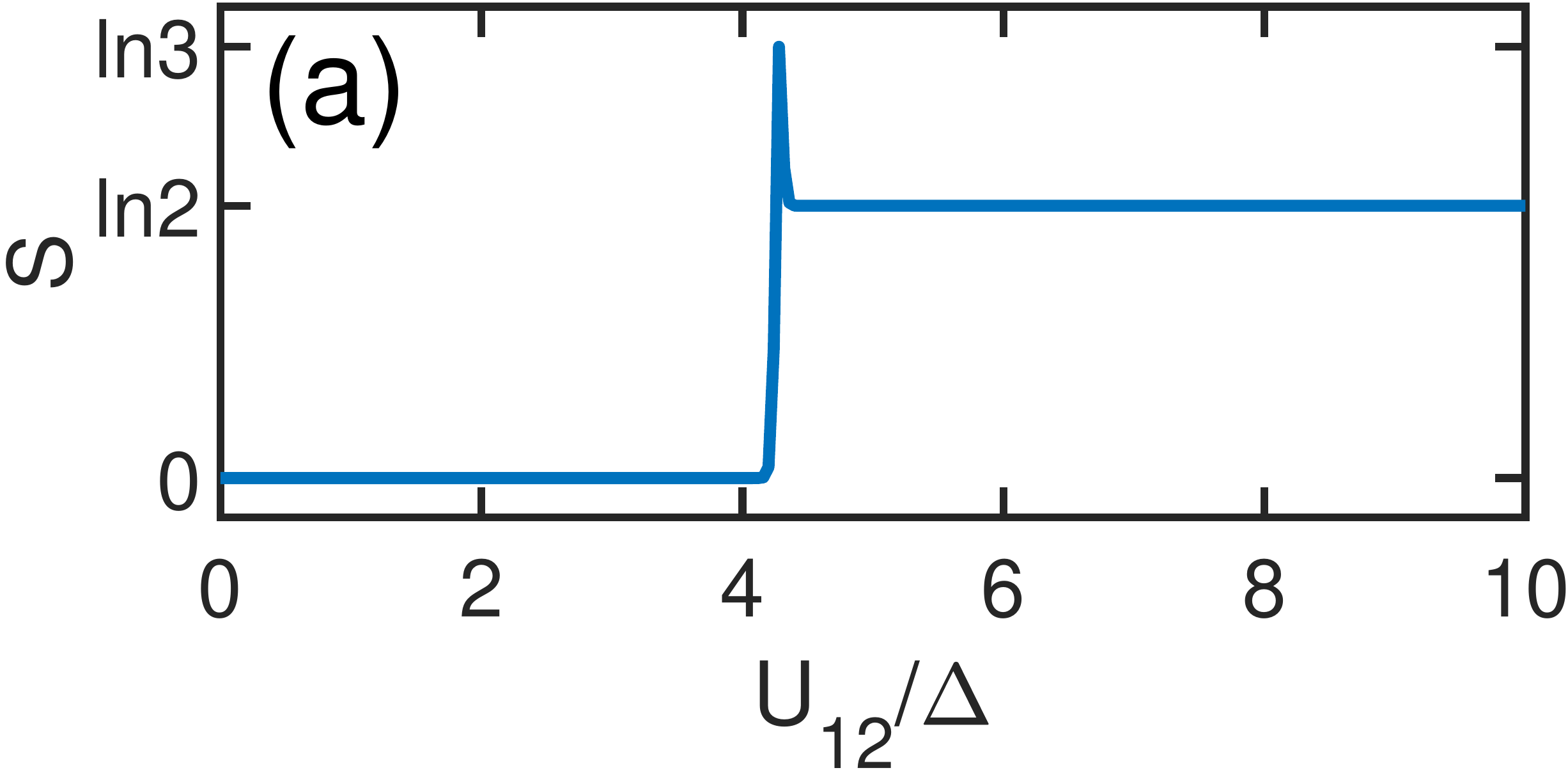}
\includegraphics[width=1\linewidth]{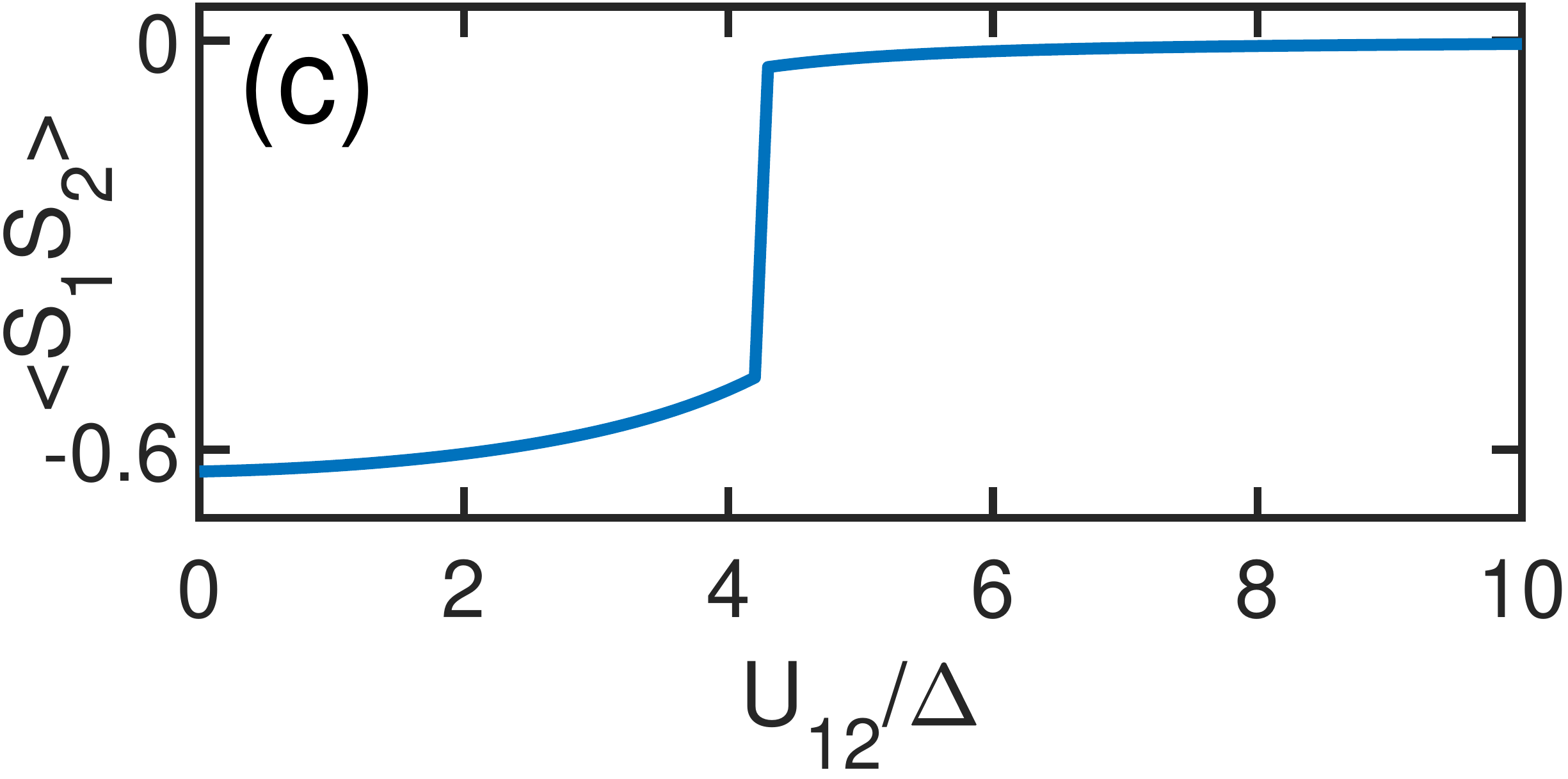}
\includegraphics[width=1\linewidth]{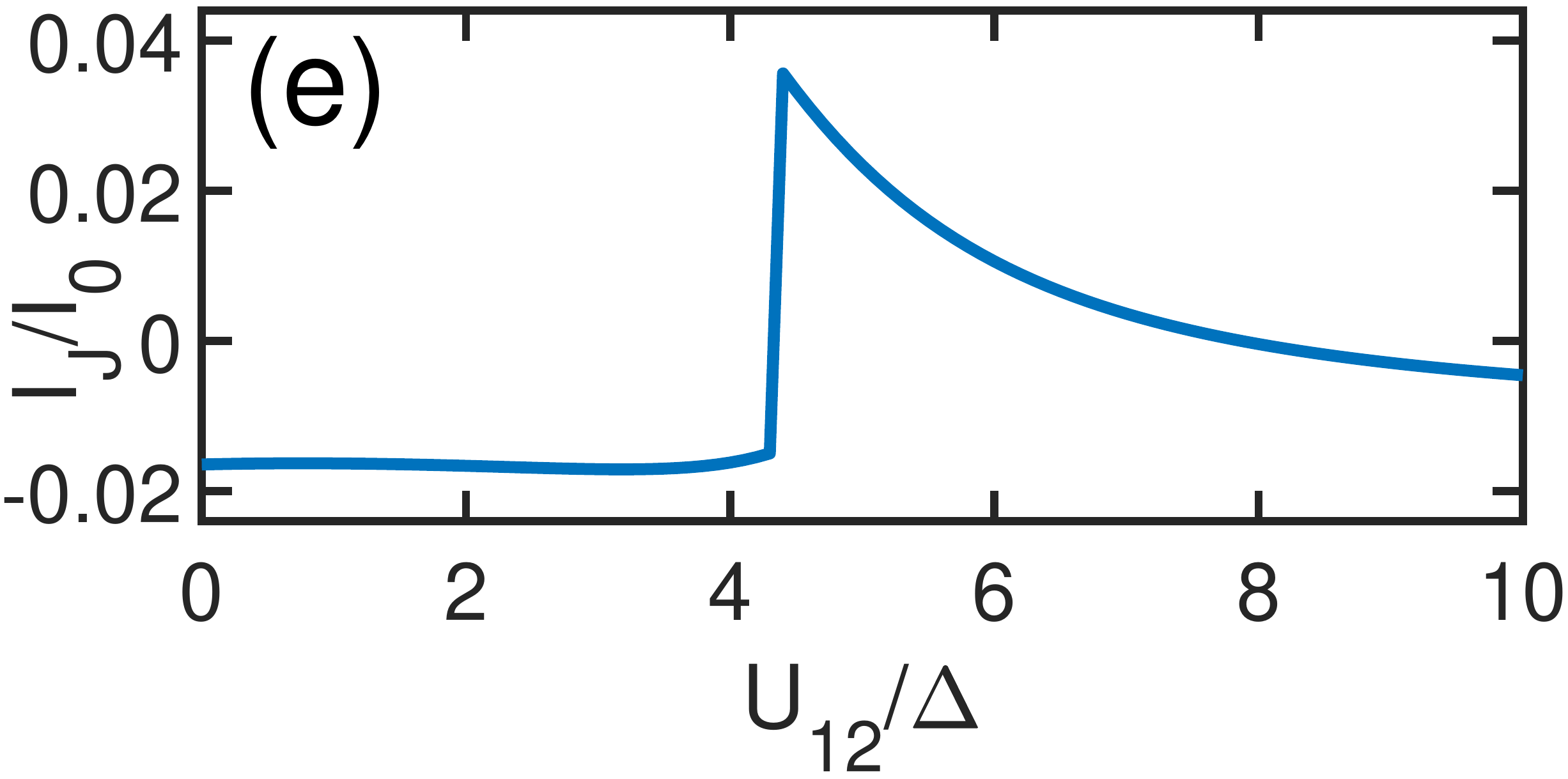}
\end{minipage}
\centering
\begin{minipage}{0.49\linewidth}
\includegraphics[width=1\linewidth]{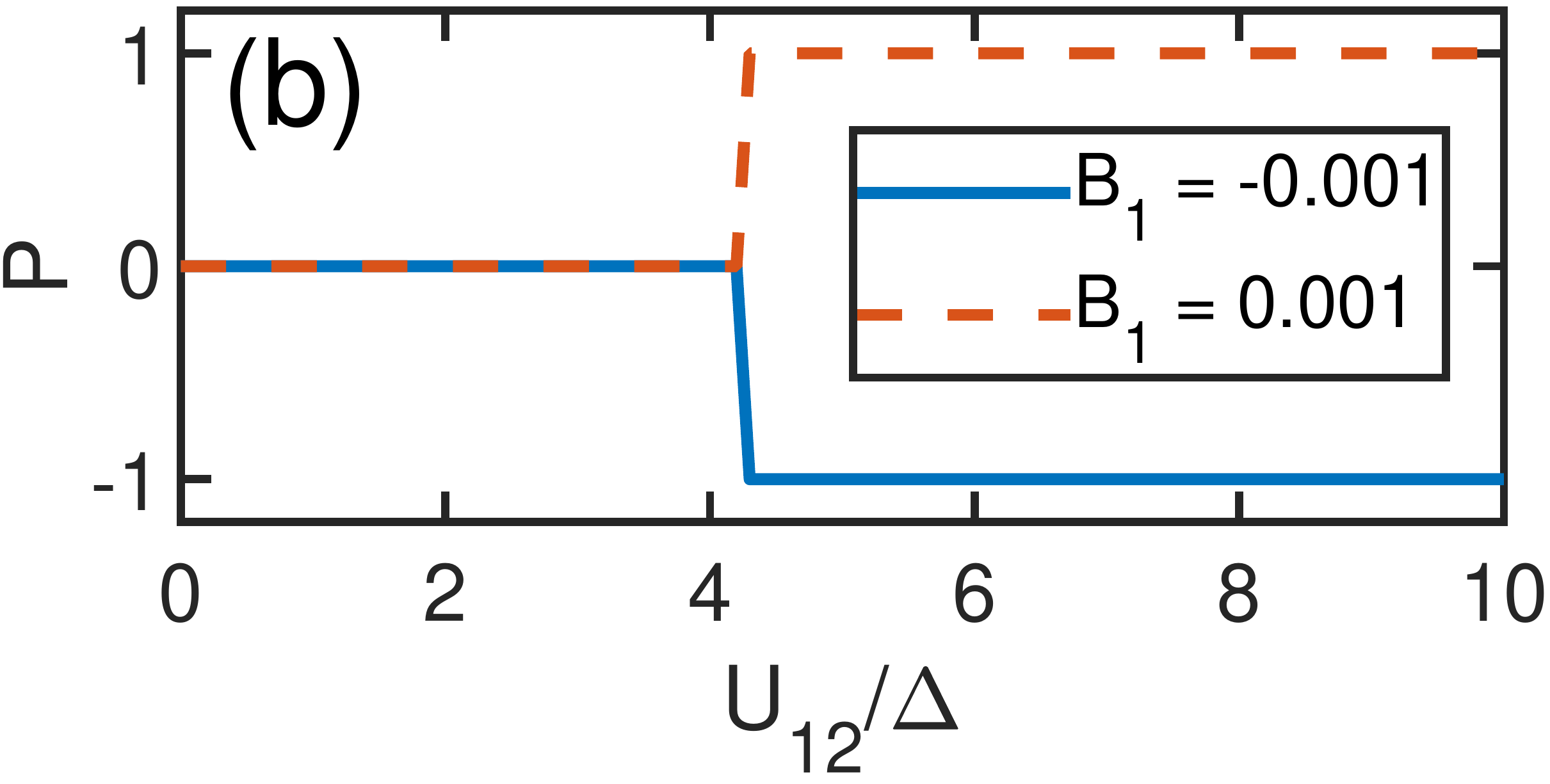}
\includegraphics[width=1\linewidth]{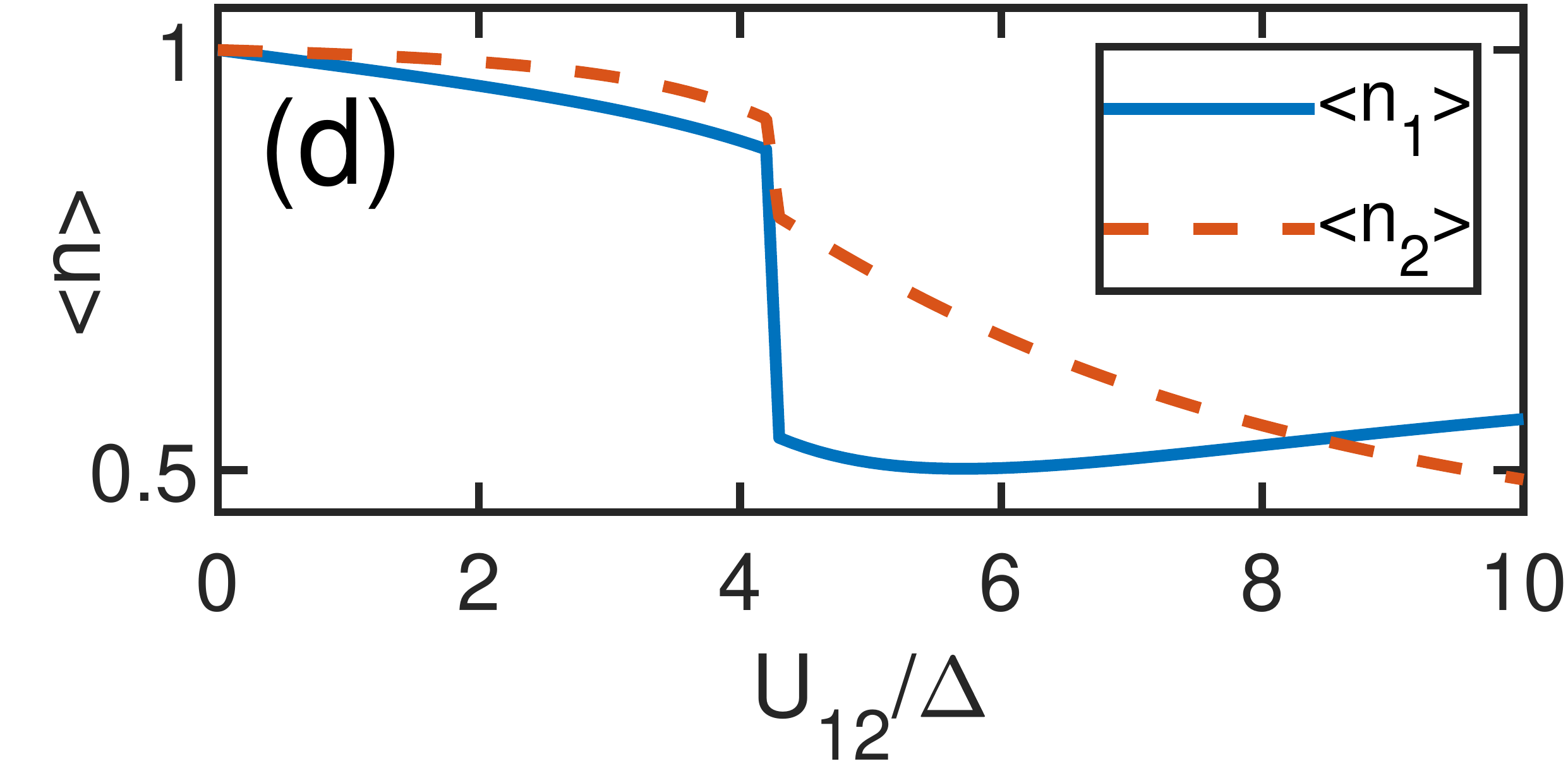}
\includegraphics[width=0.97\linewidth]{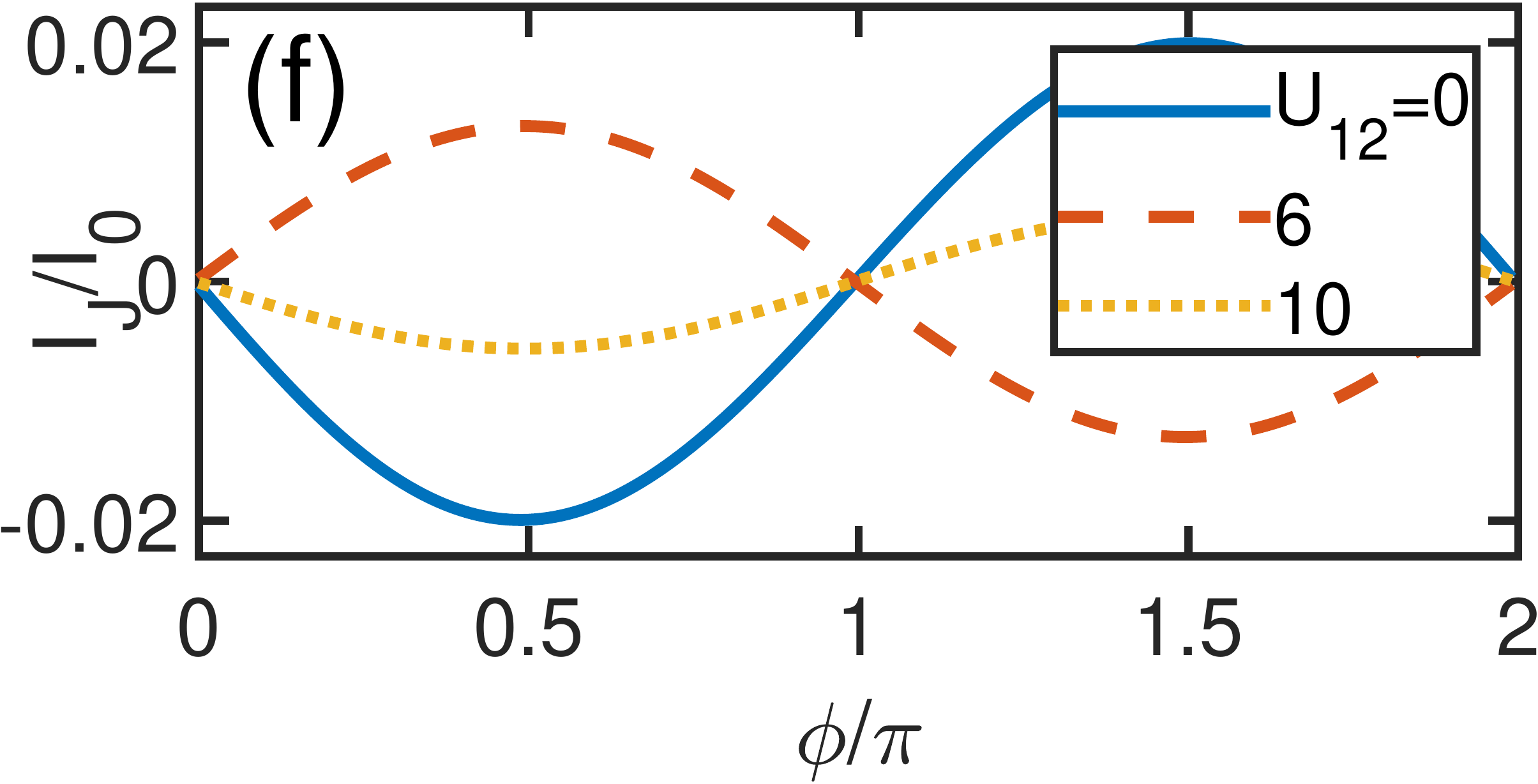}
\end{minipage}
\caption{(Colour online) The entropy with temperature is 0.005, (b) $P$ for different magnetic filed, (c) $\langle \vec{S}_{1} \!\!\cdot \!\!\vec{S}_{2} \rangle$, (d) $\langle d_{1\downarrow}d_{1\uparrow} \rangle/\Delta$ and $\langle d_{2\downarrow}d_{2\uparrow} \rangle/\Delta$ and (e) $I_{J}/I_{0}$ at $\phi=0.3\pi$ as a function of interaction between QD1 and QD2, i.e. $U_{12}/\Delta$. (f)$I_{J}/I_{0}$ versus phase difference $\phi$ for various interaction between QD1 and QD2. The interaction of QD1 and QD2 are $10\Delta$. The other parameters are given the same as in Fig. 3 unless specifically stated.}.
\end{figure}

Figure 6(a) displays the entropy $S$ versus the interdot Coulomb interaction $U_{12}$ at temperature $T=0.005$, with the fixed on-site interactions $U_{1}=U_{2}=10\Delta$, revealing a QPT at $U_{12} \approx 4.3\Delta$ where the ground state changes from a singlet (pre-QPT) to a doublet (post-QPT), as further evidenced by the parity $P$ in Fig. 6(b) transitioning from $0$ to $\pm1$ across the QPT point, while Fig. 6(c) shows the interdot spin correlation $\langle \vec{S}_{1} \!\!\cdot\!\! \vec{S}_{2} \rangle$ maintaining antiferromagnetic character ($\langle \vec{S}_{1} \!\!\cdot\!\! \vec{S}_{2} \rangle < 0$) before the QPT but vanishing ($\langle \vec{S}_{1} \!\!\cdot\!\! \vec{S}_{2} \rangle \approx 0$) afterward, and Fig. 6(d) demonstrates the occupation numbers $n_{1}$ and $n_{2}$ of both quantum dots deviating from unity at finite $U_{12}$ even at the nominal particle-hole symmetric point $\epsilon_{i} = -U_{i}/2$ due to $U_{12}$-induced particle-hole symmetry breaking, with both occupations exhibiting an abrupt decrease at the QPT point that coincides with the entropy, parity, and spin correlation transitions.

We now examine the post-transition ground state configuration of the system, where the strong interdot interaction $U_{12}$ enforces an occupancy constraint such that the two QDs collectively accommodate at most one electron, as evidenced by the $\pm 1$ parity and vanishing spin correlation $\langle \vec{S}_{1} \!\!\cdot\!\! \vec{S}_{2} \rangle$ in this regime, leading to a doublet ground state that can be expressed as the superposition $A_{3}|\!\!\uparrow_{1}\!\!0_{2}\rangle+B_{3}|0_{1}\!\!\uparrow_{2}\rangle$ or $A_{3}|\!\!\downarrow_{1}\!\!0_{2}\rangle+B_{3}|0_{1}\!\!\downarrow_{2}\rangle$, representing quantum states where one quantum dot remains empty while the other hosts a single electron with definite spin orientation, with the occupation number evolution demonstrating that increasing $U_{12}$ enhances the probability amplitude $A_{3}$ for QD1 occupation while suppressing $B_{3}$ for QD2 occupation, as reflected in the systematic growth of $n_1$ and reduction of $n_2$ with stronger interdot coupling. The behavior stems from QD2's complete isolation from the superconducting reservoir, whereby the enhanced interdot Coulomb interaction $U_{12}$ effectively blocks electron tunneling into QD2 through the mechanism of Coulomb blockade, consequently suppressing the probability amplitude $B_{3}$ while conversely amplifying $A_{3}$ due to the constraint $|A_{3}|^2 + |B_{3}|^2 = 1$ for the normalized ground state wavefunction.

Figure 6(e) presents the Josephson current $I_J$ versus the interdot interaction strength $U_{12}$ at a fixed superconducting phase difference $\phi=0.3\pi$, where a distinct discontinuity accompanied by current sign reversal is observed at the QPT point - transitioning from negative current ($I_J<0$) in the pre-QPT regime where QD1's single-particle state permits only single-electron tunneling processes , to positive current ($I_J>0$) in the post-QPT regime where QD1 enters a quantum superposition $A_{3}|0_1\sigma_2\rangle+B_{3}|\sigma_10_2\rangle$ that enables both Cooper pair tunneling and single-electron tunneling, with the Cooper pair contribution becoming dominant; moreover, the current exhibits a continuous evolution from positive to negative values with increasing $U_{12}$ in the post-QPT phase, reflecting the progressive suppression of Cooper pair tunneling (as $A_{3}$ increases and $B_{3}$ decreases) and the corresponding enhancement of single-electron tunneling, ultimately causing the current to change sign when the single-particle contribution overwhelms the Cooper pair contribution at large $U_{12}$ values.

Figure 6(f) displays the Josephson current $I_J(\phi)$ as a function of the superconducting phase difference $\phi$ for varying interdot interaction strengths $U_{12}$, revealing three distinct transport regimes: (i) in the non-interacting case ($U_{12}=0$), the current remains negative ($I_J<0$) throughout the phase range $0\leq\phi\leq\pi$, characteristic of single-electron dominated transport; (ii) at intermediate interaction ($U_{12}=6\Delta$), the current reverses sign to positive values ($I_J>0$) due to the QPT that enhances Cooper pair tunneling; and (iii) in the strong coupling regime ($U_{12}=10\Delta$), the current undergoes another sign reversal back to negative values without an accompanying phase transition, consistent with the previously discussed mechanism.

\subsection{Ferromagnetic-antiferromagnetic phase transition}
Our previous analysis has demonstrated that the system enters an antiferromagnetic state with negative spin correlation when both QD1 and QD2 experience strong on-site Coulomb interaction, though these results were obtained in the zero-field limit where the Zeeman effect is absent; the introduction of an external magnetic field $B$ may potentially induce a transition to a ferromagnetic configuration ($\langle \vec{S}_1 \!\!\cdot\!\! \vec{S}_2 \rangle > 0$) through spin polarization, which motivates our investigation in Fig. 7 of the magnetic field dependence of the quantum phase transitions.

\begin{figure}[h]
\centering
\includegraphics[scale=0.2]{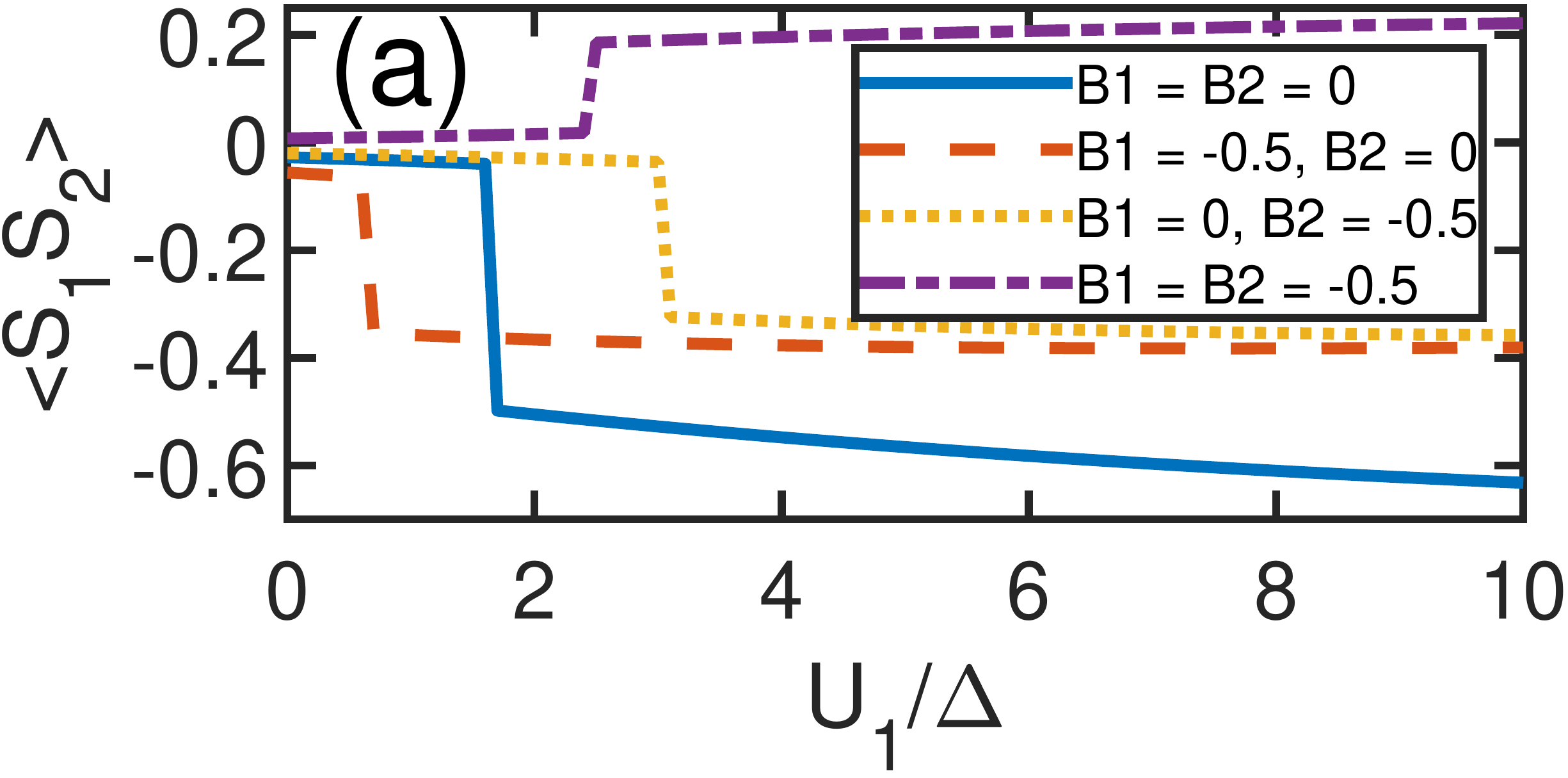}
\includegraphics[scale=0.198]{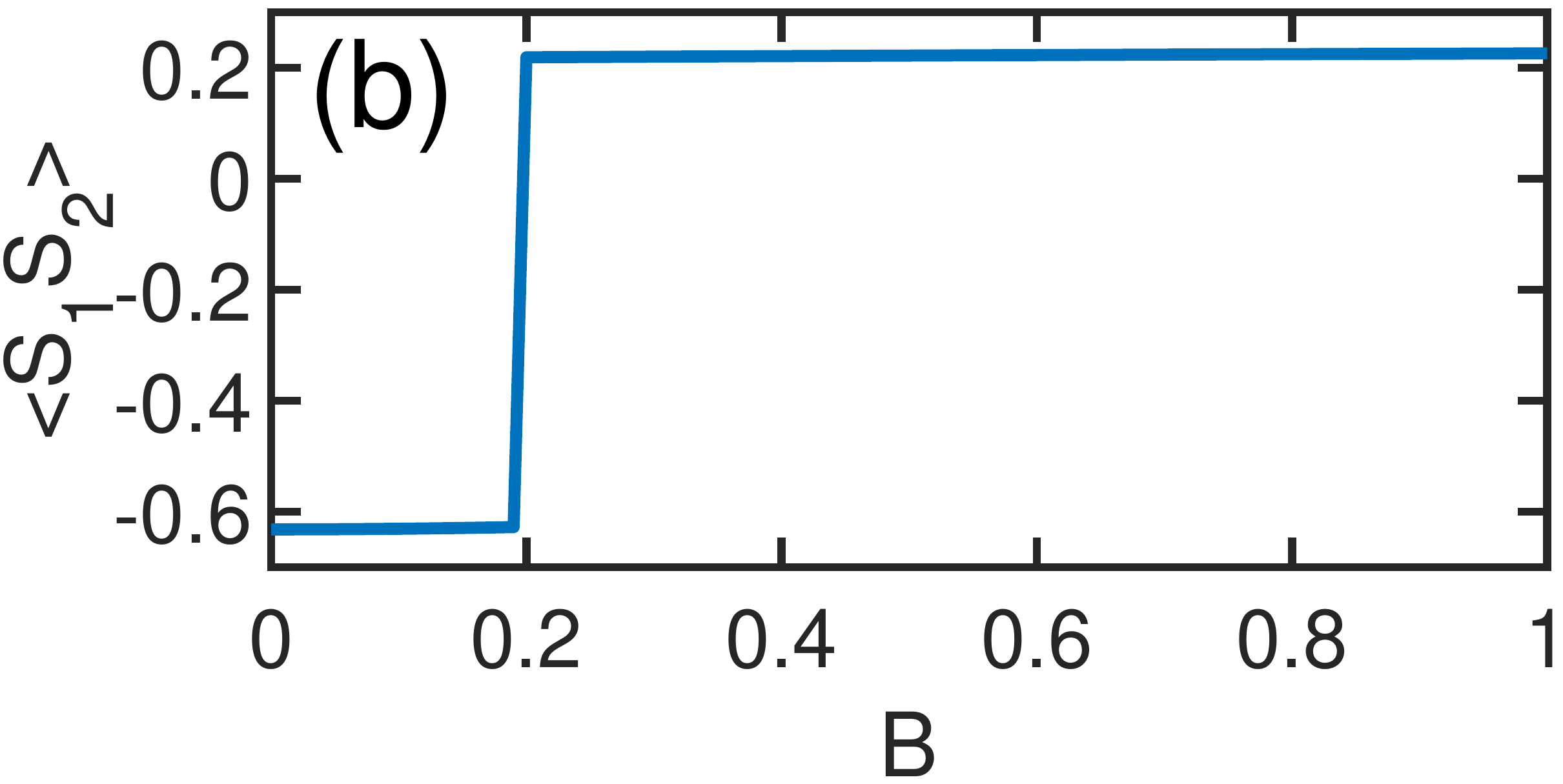}
\includegraphics[scale=0.198]{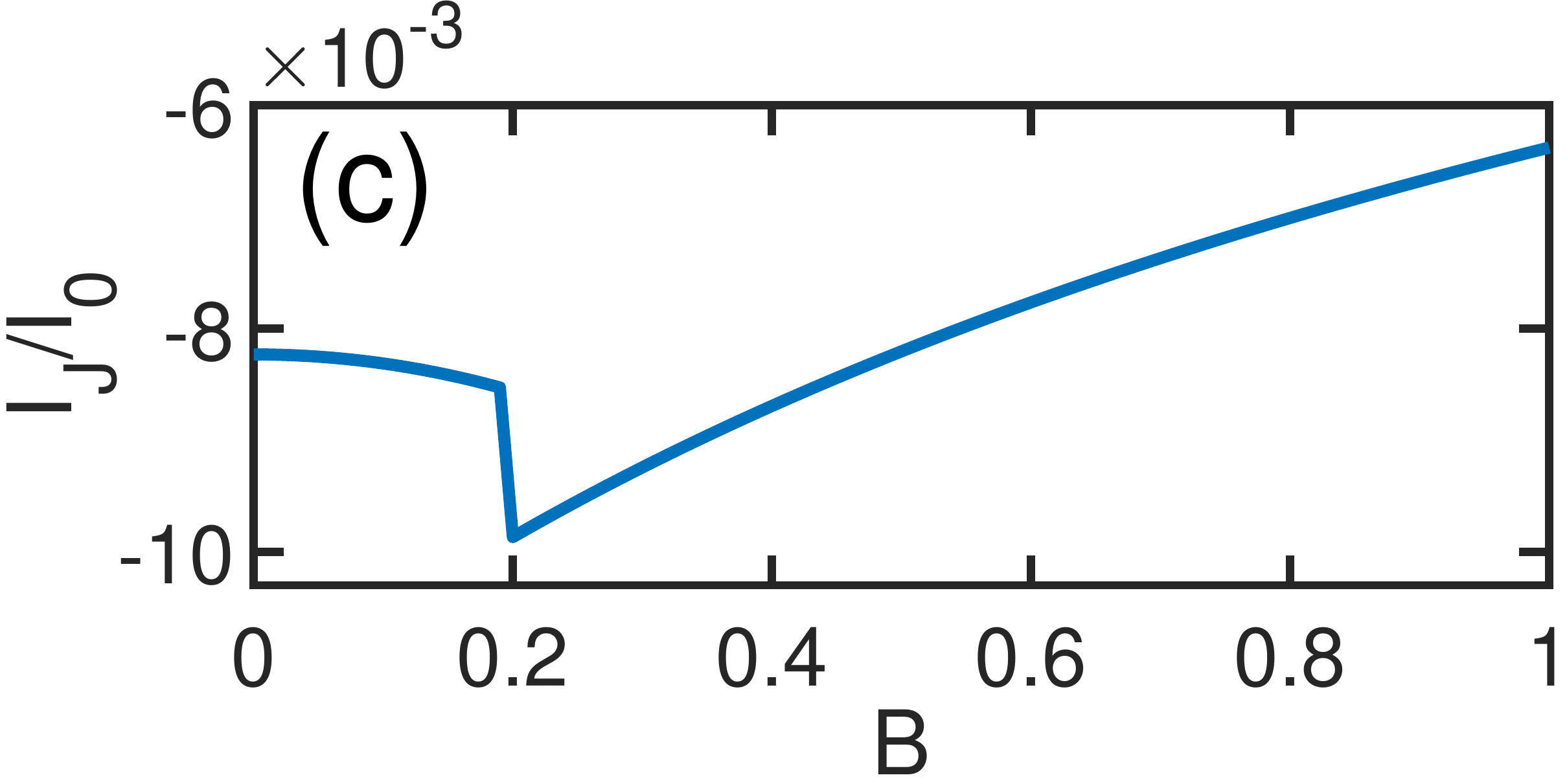}
\caption{(Colour online) (a) $\langle \vec{S}_{1} \!\!\cdot \!\!\vec{S}_{2} \rangle$ as a function of $U_{1}$ for various external magnetic fields. (b) The spin correlation as a function of external parallel magnetic field $B = -B_{1} = -B_{2}$ with $U_{1} = 10\Delta$. (c) $I_{J}/I_{0}$ as a function of external parallel magnetic field when phase difference between two SC leads is $0.3\pi$. The interaction of QD2 is $U_{2} = 10\Delta$. The other parameters are given the same as in Fig. 3 unless specifically stated.}.
\end{figure}

Figure 7(a) systematically examines the interdot spin correlation as a function of $U_1$ under various magnetic field configurations while maintaining $U_2=10\Delta$ and $U_{12}=0$, demonstrating that a magnetic field applied solely to QD1 reduces the critical interaction strength $U_{1c}$ for the QPT by disrupting the ABS in QD1 through Zeeman splitting that effectively mimics additional Coulomb interaction, whereas applying the field exclusively to QD2 increases $U_{1c}$ since the magnetic field's influence on QD2 is energetically equivalent to enhancing its local interaction strength, as corroborated by the phase diagram in Fig. 5(b) showing expanded parameter regions for the singlet phase; when co-directional magnetic fields $B_1=B_2=B$ are applied to both QDs, the post-QPT ground state transitions from antiferromagnetic ($\langle \vec{S}_1 \!\!\cdot\!\! \vec{S}_2 \rangle<0$) to ferromagnetic ($\langle \vec{S}_1 \!\!\cdot\!\! \vec{S}_2 \rangle>0$) ordering, though the precise nature of this transition under strong interactions ($U_1=U_2=10\Delta$) remained ambiguous until Fig. 7(b) revealed an abrupt change in spin correlations at a critical field strength $B_c\approx0.2$, unambiguously demonstrating the magnetic field's capability to induce a sharp QPT between these competing magnetic phases when $U_{12}=0$.

Our previous analysis has demonstrated that while interaction induced QPTs can reverse the sign of the Josephson current, the magnetic field driven ferromagnetic-antiferromagnetic transition exhibits qualitatively different transport characteristics, as evidenced in Fig. 7(c) where we plot the Josephson current $I_J$ versus parallel magnetic field $B$ at fixed phase difference $\phi=0.3\pi$, revealing that although the current shows an abrupt change at the critical field $B_c \approx 0.2$ marking the ferromagnetic-antiferromagnetic transition, it maintains a negative value ($I_J<0$) throughout the transition due to the persistent single particle nature of the transport: in the pre-transition antiferromagnetic phase ($B<B_c$) the ground state $(|\!\!\uparrow_1\downarrow_2\rangle-|\!\!\downarrow_1\uparrow_2\rangle)/\sqrt{2}$ permits only single-electron tunneling, and in the post-transition ferromagnetic phase ($B>B_c$) the aligned spin configuration $|\!\!\downarrow_1\downarrow_2\rangle$ similarly restricts current to single-particle processes, with the absence of Cooper pair tunneling in both regimes resulting in the maintained negative current direction, in stark contrast to the interaction-induced QPT case where the current sign reversal originates from the emergence of Cooper pair transport channels.

\subsection{Nonlocal magnetization}
Our prior investigations have rigorously demonstrated that the system exhibits odd parity ($P=\pm1$) under two distinct interaction regimes: (i) the $U_1=U_{12}=0$ with strong $U_2$ coupling, and (ii) the symmetric strong-coupling scenario where $U_1=U_2=10\Delta$ with large interdot interaction $U_{12}$. The application of a weak magnetic field $B_1=-0.001$ to QD1 induces Zeeman splitting of the ground state energy levels, which removes the degeneracy and drives the system into a spin-polarized configuration characterized by well-defined parity ($P=1$ or $P=-1$) and quantized total spin ($S_z=0.5$ or $S_z=-0.5$).

\begin{figure}[h]
\centering
\includegraphics[scale=0.2]{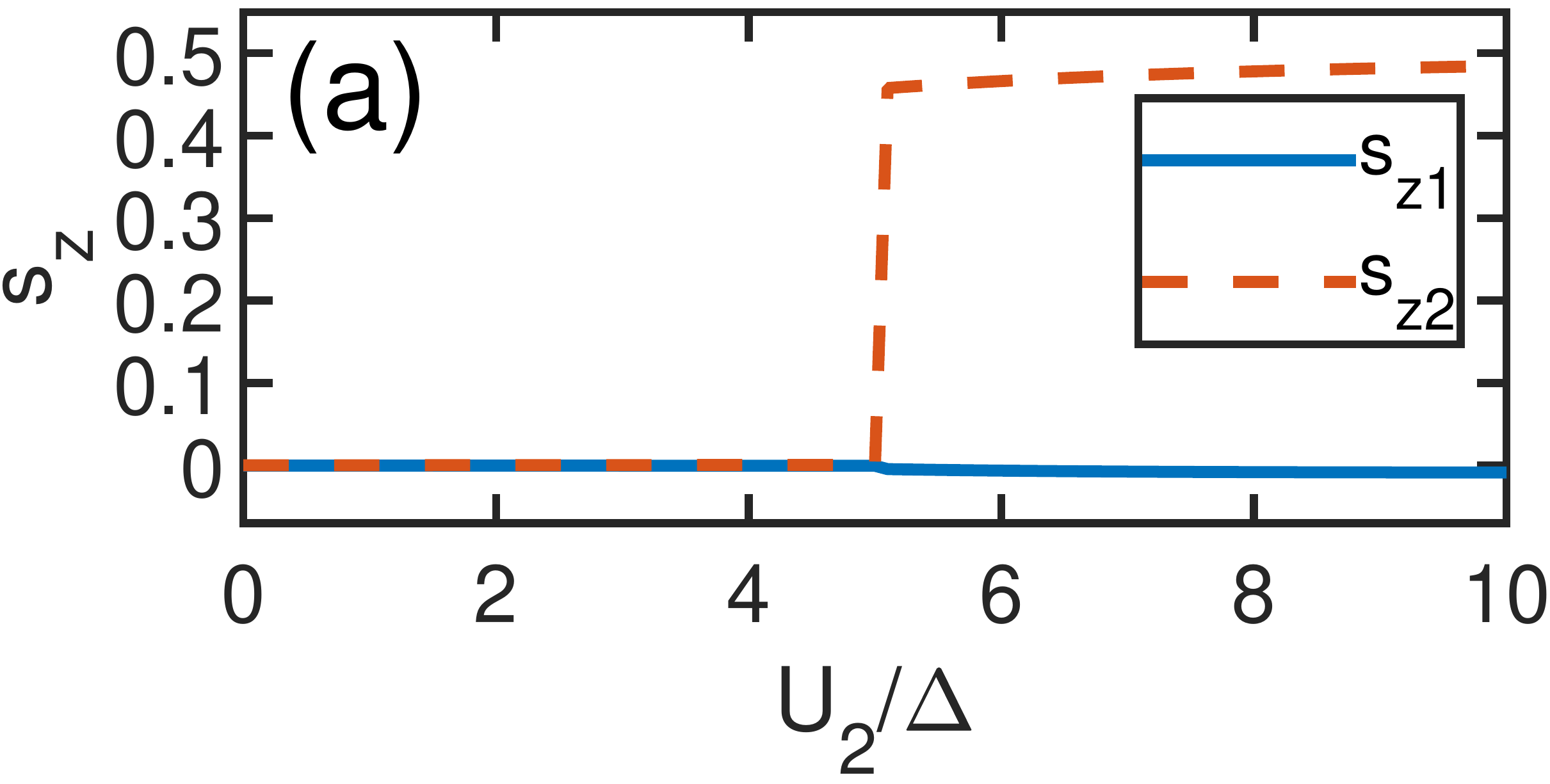}
\includegraphics[scale=0.193]{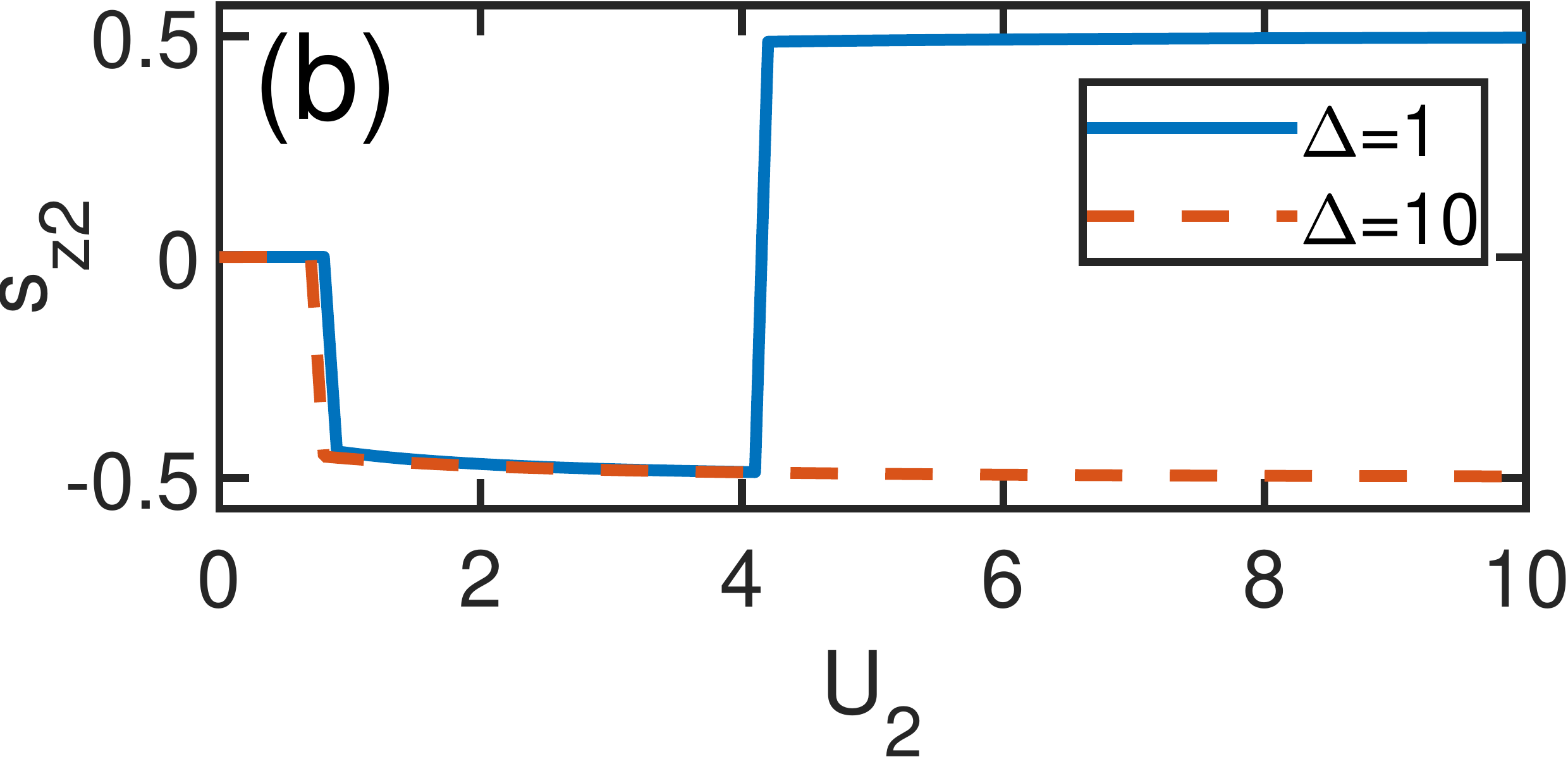}
\includegraphics[scale=0.193]{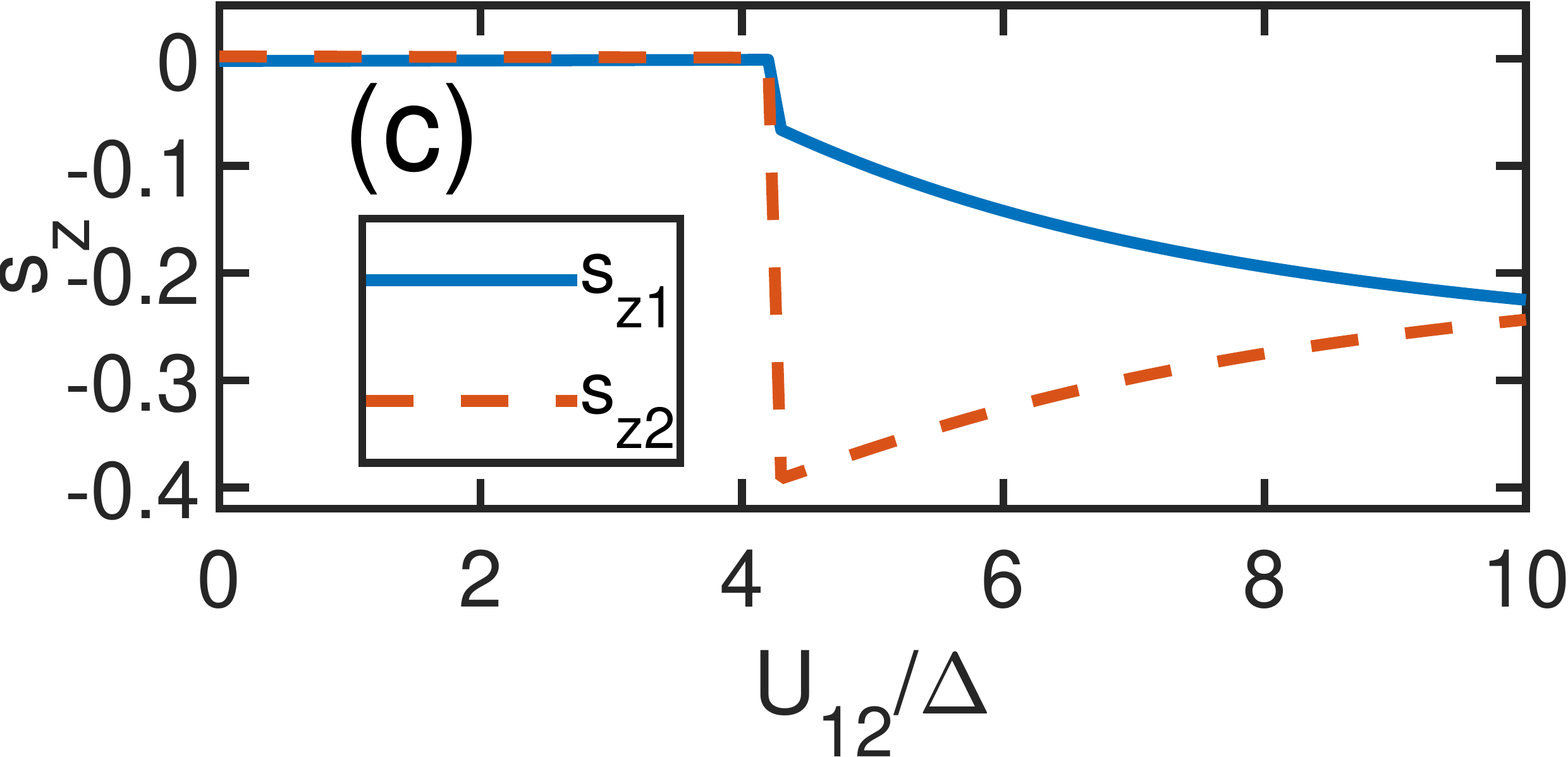}
\caption{(Colour online) (a) The magnetization intensity of QD1 and QD2, i.e. $\langle s_{z i} \rangle = 0.5(\langle n_{i\uparrow} \rangle-\langle n_{i\downarrow} \rangle)$, as a function of $U_{2}$ with $U_{1} = 0$. (b) $\langle s_{z 2} \rangle$ as a function of $U_{2}$ for varying superconductor gap $\Delta$ with $U_{1} = 0$. (c) $\langle s_{z 1} \rangle$ and $\langle s_{z 2} \rangle$ as a function of $U_{12}$ with $U_{1} = U_{2} = 10\Delta$. The other parameters are given the same as in Fig. 3 unless specifically stated.}.
\end{figure}

Figure 8(a) examines the magnetization profiles of both quantum dots as functions of $U_{2}$ under a weak magnetic field $B_1=-0.001$ applied solely to QD1 while maintaining $U_{1}=U_{12}=0$, showing that in the pre-QPT regime both QDs exhibit zero magnetization ($\langle S_{1z}\rangle=\langle S_{2z}\rangle=0$), whereas post-QPT QD1 remains in a non-magnetic Andreev bound state ($\langle S_{1z}\rangle\approx0$) while QD2 develops nearly full magnetization ($\langle S_{2z}\rangle\approx 0.5$) as it enters a single-particle state, demonstrating remarkable non-local spin polarization mediated through the interdot coupling.

Furthermore, this figure demonstrates that QD2's magnetization develops in antiparallel alignment with the applied magnetic field direction - a striking contrast to single-dot Josephson junctions where magnetic fields induce parallel magnetization in $\pi$-phase quantum dots. To elucidate this counterintuitive behavior, Fig. 8(b) presents QD2's magnetization evolution as a function of interaction strength at $t=0.5$, comparing $\Delta=1$ (blue solid line) and $\Delta=10$ (red dashed line) cases. The $\Delta=1$ scenario shows that increasing QD2's interaction strength beyond the phase transition point drives a complete magnetization reversal from negative to positive values.

Remarkably, when applying a magnetic field exclusively to QD1 (in the ABS state), the induced magnetization emerges in QD2, demonstrating that QD2's magnetization fundamentally requires first-order electronic processes mediated through QD1. Figure. 9 delineates the two essential first-order transition mechanisms: (a) inter-QD transitions between QD1-QD2 requiring $U_{2}/2$-scaled energy at the particle-hole symmetric point ($\varepsilon_{2\sigma}=-U_{2}/2$), and (b) electron exchange between the superconducting reservoir and QD1, demanding $\Delta$-dependent energy to excite individual electrons. The system displays two characteristic regimes: for small $U_{2}$, electron transport occurs predominantly between QD1-QD2, whereas large $U_{2}$ promotes reservoir-QD1 transitions. The spin-polarizing magnetic field on QD1 enforces spin-down polarization in QD1 following first-order transitions. In systems with small $U_{2}$ values, electron tunneling predominantly occurs between QD1 and QD2, where before tunneling QD1 acquires spin-down polarization and QD2 starts in a single-electron configuration with its spin initially pointing downward, leading to magnetization that aligns parallel with the external magnetic field. On the other hand, when $U_{2}$ is large and electrons primarily transfer between the superconducting reservoir and QD1 (while QD1 still achieves spin-down polarization and QD2 remains in its single-electron state), after tunneling the nonlocal nature of the superconducting proximity effect facilitates Cooper pair formation across QD1 and QD2, thereby reversing QD2's spin to an upward orientation and generating magnetization that is oriented antiparallel to the applied magnetic field. The $\Delta=10$ scenario (red dashed line) validates this picture through pure QD1-QD2 transitions and resulting spin-down magnetization. Additionally, the ground state evolution from $|0_{1}\!\!\downarrow_{2}\rangle-|d_{1}\!\!\downarrow_{2}\rangle$ to $|0_{1}\!\!\uparrow_{2}\rangle-|d_{1}\!\!\uparrow_{2}\rangle$ with increasing $U_{2}$ potentially indicates a quantum phase transition.
\begin{figure}[H]
\centering
\includegraphics[scale=0.28]{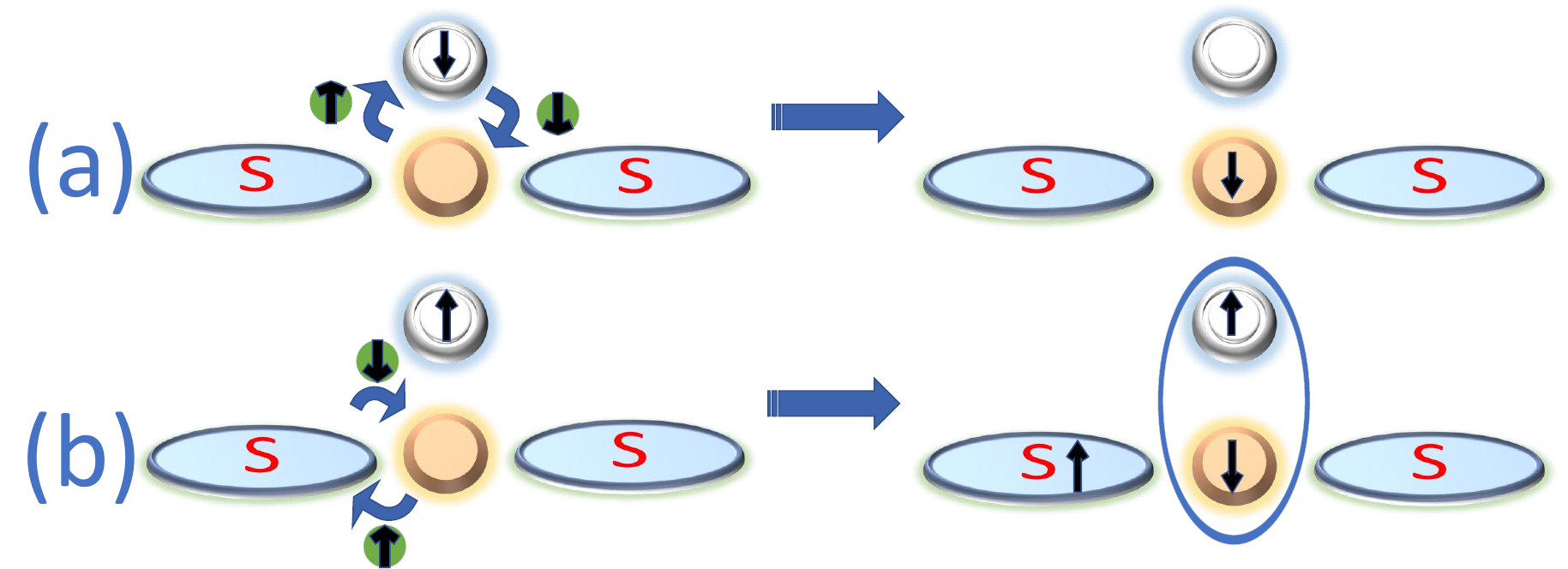}
\caption{(Color online) Schematic diagram of first-order electronic processes mediated through QD1 with $U_{1}=U_{12}=0$ and $B=-0.001$.}
\end{figure}

Figure 8(c) then investigates the $U_{12}$-dependent magnetization under identical field conditions but with strong on-site interactions $U_{1}=U_{2}=10\Delta$, where both QDs initially show zero magnetization before the QPT but acquire finite spin polarization afterward following the relation $\langle S_{1z}\rangle+\langle S_{2z}\rangle\approx -0.5$.

In this configuration, both quantum dots exhibit magnetization parallel to the applied magnetic field direction, as evidenced by the field-free ground state being a superposition of $|\!\!\downarrow_{1}\!\!0_{2}\rangle+|0_{1}\!\!\downarrow_{2}\rangle$ or $|\!\!\uparrow_{1}\!\!0_{2}\rangle+|0_{1}\!\!\uparrow_{2}\rangle$ states (with QD1 in a single-electron configuration), where the introduction of a downward-oriented magnetic field induces spin polarization in QD1, energetically favoring the $|\!\!\downarrow_{1}0_{2}\rangle+|0_{1}\!\!\downarrow_{2}\rangle$ state as the ground state and consequently causing both QD1 and QD2 to develop magnetization that is parallel to the external field direction.

Notably, the post-transition magnetization exhibits a distinct $U_{12}$-dependent characteristic: the magnetization intensity of QD1 gradually increases as $U_{12}$ increases, whereas QD2's magnetization intensity shows a corresponding decreasing. This behavior can be quantitatively explained by analyzing the post-transition ground state wavefunction $\psi = A_{3}|\!\!\downarrow_{1}\!\!0_{2}\rangle+B_{3}|0_{1}\!\!\downarrow_{2}\rangle$, where the coefficient $A_{3}$ systematically increases while $B_{3}$ decreases with growing $U_{12}$. Such variation in coefficients results in strengthened single-particle occupation in QD1 and weakened occupation in QD2, thereby providing a direct microscopic explanation for the observed magnetization evolution. In these figures, the combined magnetization of QD1 and QD2 does not precisely sum to $\pm 0.5$, as the superconducting reservoirs introduce a minor but non-negligible magnetization contribution.

\section{Conclusion}
In this work, we examine a Josephson junction system comprising two coupled QDs connected to two superconducting leads. Utilizing a discretization technique, we approximate the self-energy of the superconducting leads by replacing them with a finite set of discrete sites. The effective Hamiltonian of the system is then solved exactly through state-space expansion-based diagonalization. We systematically calculate key physical quantities including the system entropy, parity, inter-dot spin correlation, induced superconducting order within the QDs, QD occupation numbers, and Josephson current across diverse parameter regimes.

Our results demonstrate that with vanishing on-site interaction in QD1 ($U_{1} = 0$) and zero interdot interaction ($U_{12} = 0$), modulating QD2's interaction strength triggers a QPT accompanied by discontinuous changes in all physical observables due to ground state reconstruction. Remarkably, we identify a second QPT when QD2's on-site interaction exceeds a critical threshold (while maintaining $U_{12} = 0$) and QD1's interaction is varied. Furthermore, we reveal a third distinct QPT pathway when both QDs possess strong on-site repulsion and the interdot interaction $U_{12}$ is varied.

Our investigation further explores how external magnetic fields influence the system's antiferromagnetic ground state, revealing that magnetic fields applied to either QD1 or QD2 individually modify the phase transition boundaries. Notably, when parallel magnetic fields are simultaneously applied to both quantum dots, the spin correlation undergoes a sign reversal from negative to positive values, signaling a transformation from antiferromagnetic to ferromagnetic ordering. Importantly, despite this magnetic-field-induced quantum phase transition, the fundamental direction of the Josephson current remains unchanged.

Our study further examines how a weak magnetic field applied exclusively to QD1 influences the system's magnetization characteristics, revealing that in odd-parity ground states, such localized magnetic perturbation induces a striking non-local magnetization response where QD2 unexpectedly develops magnetization despite the field being applied only to QD1. Moreover, we establish that the orientation of nonlocal magnetization on QD2 can be controllably reversed through precise adjustment of the interaction strength $U_{2}$ under the condition where both $U_{1}$ and $U_{12}$ are set to zero.

\bibliography{ref}

\end{document}